\documentclass[journal]{IEEEtran}
\usepackage{amsmath}  
\usepackage{graphicx}
\usepackage[hidelinks]{hyperref}
\usepackage{cite}

\usepackage{picinpar}
\usepackage{url}
\usepackage{flushend}
\usepackage[latin1]{inputenc}
\usepackage{colortbl}
\usepackage{soul}
\usepackage{multirow}
\usepackage{pifont}
\usepackage{color}
\usepackage{alltt}
\usepackage{enumerate}
\usepackage{siunitx}
\usepackage{epstopdf}
\usepackage{pbox}

\begin{document}

\title{Frequency Response and Eddy Current Power Loss in Magneto-Mechanical Transmitters}

 \author{Jiheng~Jing,
        Sameh~Tawfick,
        and~Gaurav~Bahl%
\thanks{J. Jing, S. Tawfick, and G. Bahl are affiliated with the Department of Mechanical Science and Engineering, University of Illinois at Urbana-Champaign, Urbana, IL 61801 USA (e-mail: bahl@illinois.edu).}%
}

\maketitle

\begin{abstract}
Magneto-mechanical transmitters offer a compact and low-power solution for the generation of ultra-low frequency (ULF) magnetic signals for through-ground and through-seawater communications. Resonant arrays of smaller magneto-mechanical transmitters are particularly interesting in this context as the physical scaling laws allow for the increase of operating frequency and reduce the power requirements for ULF signal generation. In this work, we introduce a generalized model for accurate prediction of frequency and mode shape in generalized magneto-mechanical resonator arrays (MMRAs) that accounts for near-field magnetic interactions as well as magnetically induced nonlinearity. Using experiments, we demonstrate that our predictive capability is significantly improved compared against simplified dipole approximations. We additionally model the eddy current losses internal to the array and find that they are in agreement with experimental observations.
\end{abstract}


\IEEEpeerreviewmaketitle

\section{Introduction}

\IEEEPARstart{R}{adio} frequency communications are among the most important technologies today and are deployed in an extremely wide range of applications. Unfortunately, RF communications cannot be used in conductive environments, e.g., for undersea and through-ground transmission, due to the extremely low skin depth at frequencies above a few kHz. On the other hand, electromagnetic signals at ultra-low frequencies (ULF, 0.3-3 kHz) exhibit significantly greater skin depth and are therefore used for low-bandwidth underground communication via inductive coils \cite{5452976} and for submarine communications at great technical expense \cite{Wait1972, Gigliotti2009, alma99955070279905899}. The biggest challenge with ULF systems is the significant power requirement \cite{Wait1972, Gigliotti2009} and the need for large antenna structures that cannot be easily transported \cite{alma99955070279905899}. Recently, an innovative approach has emerged for the ULF transmitter challenge \cite{Rhinithaa, UCLA, Gong2018, 8304977, 8917925,8457248,8939554, esmm5f497}, in which the large magnetic field is produced by a fixed magnetic dipole, while the ULF carrier frequency generation is achieved through either torsional mechanical oscillation or rotation of this dipole. It has been shown that this approach can reduce the transmitter power requirements from a few megawatts to tens of watts, and the transmitter size from a few kilometers to tens of centimeters \cite{Rhinithaa, UCLA}. Similar promising solutions using magneto-mechanics have also been proposed for improvements in ULF receivers \cite{cook_dominiak_widmer_2009, 8234677, 9672672}.

Interestingly, the scaling laws for such magneto-mechanical systems reveal that, instead of using one large magneto-mechanical oscillator, subdivision of the magnetic dipole to an array of smaller oscillators provides a major scaling advantage for increasing frequency and for reducing the power requirements \cite{Rhinithaa, UCLA}. At the same time, the technical complexity of the system does increase since we now must consider the mutual magnetic interactions within the array. Even though there is an increase in the degrees of freedom, only the mechanically synchronized mode (i.e., the in-phase mode) is of interest since it has the highest resonant frequency and produces the largest magnetic signal \cite{Rhinithaa, UCLA}. Presently, however, the models for frequency prediction and power consumption for magneto-mechanical resonators are not sufficiently advanced, especially in the context of the complexity added by mutual interactions in arrays, near-field effects, eddy current loss, and nonlinearity. 

In this paper, we generalize this problem for a magneto-mechanical resonator array (MMRA) that is magnetically actuated by a single drive coil. We present a general dynamical model for MMRAs that accounts for the near-field magnetic interactions between the oscillator elements. This model also reveals, quite surprisingly, that the commonly used coil drive mechanism is responsible for additional interaction terms between the array elements that have not previously been modeled. Using this, we demonstrate that the model's frequency prediction capability is significantly improved in both linear and nonlinear regimes. We additionally develop a simplified analytical model for estimating the eddy current loss in MMRAs, without reliance on complex finite element simulations, and verify the numerical calculations by making a comparison to experimental measurements.

\section{Dynamical model of an MMRA}
\label{sec:DynamicalModel}

\begin{figure} 
    \centering
  {%
       \includegraphics[width=\linewidth]{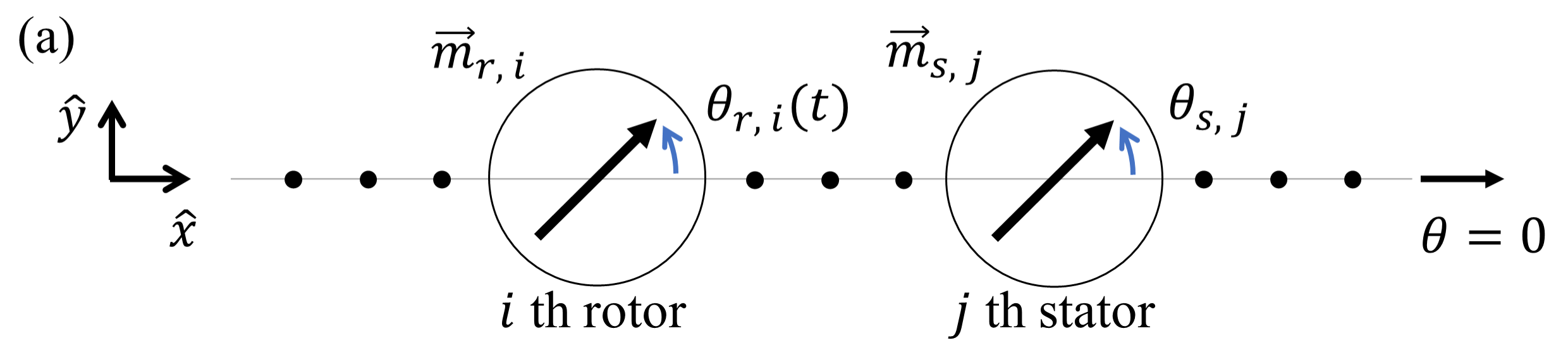}}
\\
  {%
        \includegraphics[width=\linewidth]{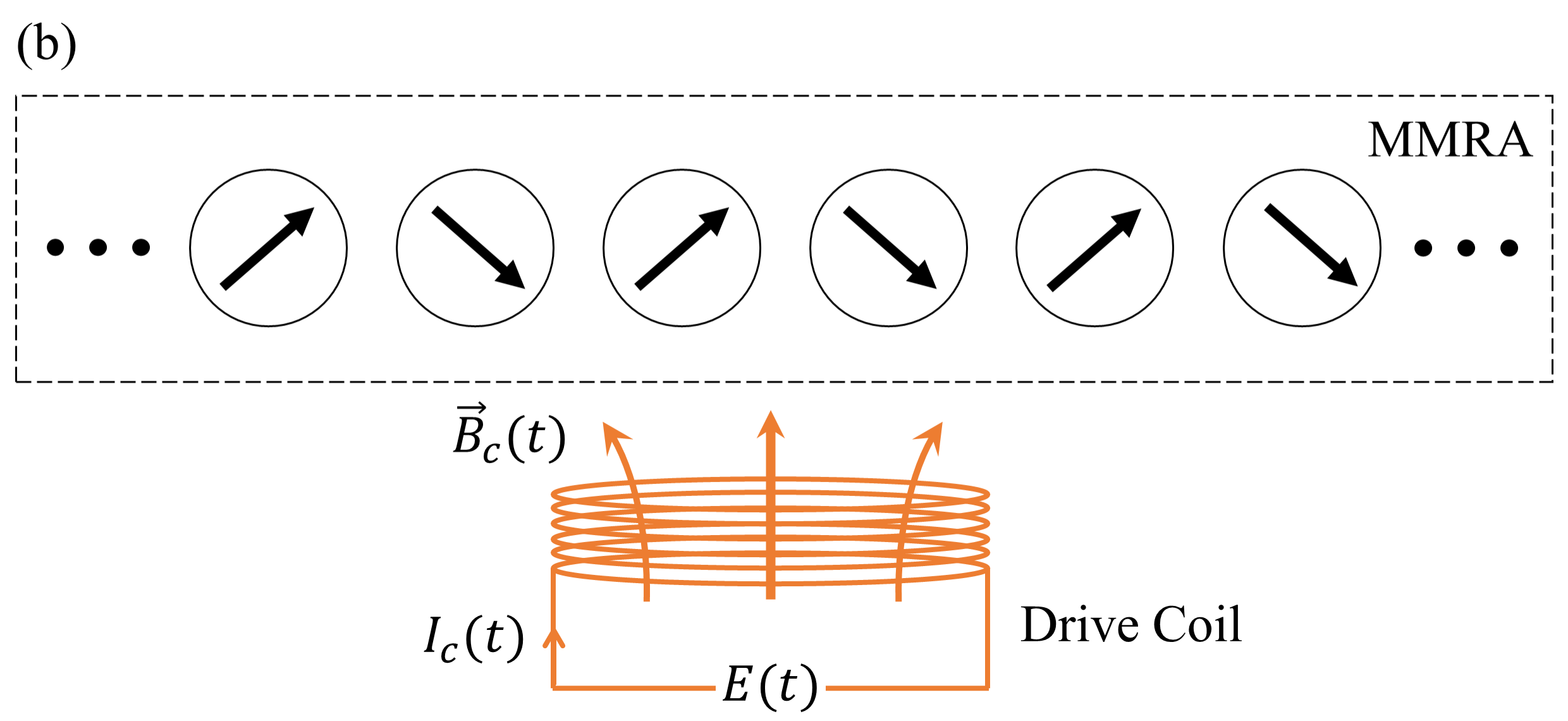}}
        \caption{\textbf{(a)} Schematic of an MMRA consisting of an arbitrary linear chain of magnetic rotors and stators. Rotors are rotating about their longitudinal axes and stators are fixed at certain angles. $\theta_{r,\,i}(t)$ represents the time-varying angular position of the $i^\textrm{th}$ rotor. $\theta_{s,\,j}$ represents the constant angular position of the $j^\textrm{th}$ stator. $\vec{m}_{r,\,i}$ and $\vec{m}_{s,\,j}$ represent the magnetic moments of the $i^\textrm{th}$ rotor and $j^\textrm{th}$ stator, respectively. $\theta = 0$ is defined along the $\hat{x}$ axis. \textbf{(b)} Schematic of an MMRA coupled to a drive coil. The drive coil generates a time-varying magnetic field $\vec{B}_c(t)$. The oscillation of rotors generates a time-varying back emf $E(t)$ across the coil, and thus a time-varying current $I_c(t)$ flowing through the coil.}
  \label{System} 
\end{figure}

We consider a magneto-mechanical system consisting of a linear chain of magnetized torsional mechanical resonators (rotors), interspersed with magnetized stator elements at fixed angles, all having uniform magnetization perpendicular to the $\hat{z}$-direction [Fig.~\ref{System}(a)]. The equation of motion for the $n^\textrm{th}$ rotor is given by the second order ordinary differential equation (in the undamped form):

\begin{align}\label{eom_rotor_general}
    J_n \ddot{\theta}_{r,\,n} = \tau_{\textrm{sus},\, n} + \sum\limits^{\substack{\textrm{all} \\ \textrm{rotors}}}_{\substack{i \\ i \neq n}} \tau_{r,\,ni} + \sum\limits^{\substack{\textrm{all} \\ \textrm{stators}}}_{j} \tau_{s,\,nj} + \tau_{\textrm{ext},\, n}
\end{align}
where $J_n$ is the moment of inertia of the rotor, $\theta_{r,\,n}$ is the instantaneous angle relative to the $\hat{x}$ axis [see Fig.~\ref{System}(a)], $\tau_{\textrm{sus},\, n}$ is the restoring torque associated with the mechanical suspension, $\tau_{r,\,ni}$ and $\tau_{s,\,nj}$ are the magnetic torques generated due to interaction with a different $i^\textrm{th}$ rotor and the $j^\textrm{th}$ stator respectively, and $\tau_{\textrm{ext},\, n}$ is the torque generated by external sources (e.g., the drive coil). In this work, we assume that the suspension has linear characteristics, and therefore $\tau_{\textrm{sus},\, n} = - \kappa_{\textrm{sus},\,n} \, \theta_{r,\,n}$, where $\kappa_{\textrm{sus},\,n}$ is the torsional stiffness constant. 

The simplest expressions for the magnetically-induced torques $\tau_{r,\,ni}$ and $\tau_{s,\,nj}$ can be obtained by approximating each magnetic moment as a point dipole. Using this ``dipole model'', we derive [see (\ref{Appendix_tau_dipole}) in Appendix \ref{sec:Magnetic_Torque}] the interaction torque
\begin{align}\label{tau_dipole}
    \nonumber {\tau}_{r,\,ni} = - \frac{\mu_0 m_{r,\,i} m_{r,\,n}}{4\pi |d_{ni}|^3} &\left[2\cos(\theta_{r,\,i})\sin(\theta_{r,\,n}) \right. \\ 
    &\left. +  \cos(\theta_{r,\,n})\sin(\theta_{r,\,i})\right]
\end{align}
where $\mu_0 = 4\pi\times 10^{-7}$ H/m is the vacuum permeability, $m_{r,\,i}$ and $m_{r,\,n}$ are the magnitudes of the magnetic dipole moments of the $i^\textrm{th}$ and $n^\textrm{th}$ rotors, while $d_{ni}$ is the center-to-center distance between the dipoles. This dipole model is often used \cite{Rhinithaa,UCLA,Grinberg2019,Grinberg2018} to study the resonances of magneto-mechanical systems. However, as the distance between the rotors decreases, this model becomes a poor predictor of the MMRA dynamics due to near field effects. As we show later, the discrepancy between the frequency prediction and measured resonance becomes quite significant for most magneto-mechanical transmitters \cite{Rhinithaa,UCLA} that have compactness as a design goal.

In order to better predict the magnetic interactions, we develop a new model (details in Appendix \ref{sec:Magnetic_Torque}) that considers the complete volume of each interacting magnetic element in the system. We find that this approach is far better at accounting for near field effects and results in a more accurate frequency prediction. In the most general form, and without any small angle approximation, the interaction torque $\tau_{r,\,ni}$ (generated by the $i^\textrm{th}$ rotor on the $n^\textrm{th}$ rotor) can be written as (see Appendix \ref{sec:Magnetic_Torque}):
\begin{align}\label{tau_r_general}
    \nonumber {\tau}_{r,\,ni}  = - &\left[ \kappa^{(1)}_{r,\,ni} \cos(\theta_{r,\,i}+\theta_{r,\,n}) + \kappa^{(2)}_{r,\,ni} \sin(\theta_{r,\,i})\cos(\theta_{r,\,n})\right. \\ 
    &  \left.  + \kappa^{(3)}_{r,\,ni} \cos(\theta_{r,\,i})\sin(\theta_{r,\,n}) \right]
\end{align}
where $\kappa^{(1)}_{r,\,ni}$, $\kappa^{(2)}_{r,\,ni}$, and $\kappa^{(3)}_{r,\,ni}$ are the torsional stiffness coefficients. These stiffness coefficients are all functions of the instantaneous angle $\theta_{r,\,i}$ and $\theta_{r,\,n}$, and further depend on the relative geometry and individual cross-section of the interacting rotors. Presently, we can simplify our analysis using the assumption that rotors are cylindrical with a circular cross-section, and find that $\kappa^{(2)}_{r,\,ni}$ and $\kappa^{(3)}_{r,\,ni} $ become independent of $\theta_{r,\,i}$ or $\theta_{r,\,n}$, and $\kappa^{(1)}_{r,\,ni} $ becomes zero [see (\ref{Appendix_C1_to_C4}) and (\ref{Appendix_k1_to_k3}) in Appendix \ref{sec:Magnetic_Torque}]. Similarly, the general form of the interaction torque $\tau_{s,\,nj}$ can be expressed as 
\begin{align}
    \tau_{s,\,nj} = - & \left[ \kappa^{(1)}_{s,\,nj} \cos(\theta_{s,\,j} + \theta_{r,\,n}) + \kappa^{(2)}_{s,\,nj} \sin(\theta_{s,\,j}) \cos (\theta_{r,\,n})\right. \notag \\
    & \left.  + \kappa^{(3)}_{s,\,nj} \cos(\theta_{s,\,j})\sin(\theta_{r,\,n}) \right] \label{tau_s_general}
\end{align}
where $\theta_{s,\,j}$ is the angle of the $j^\textrm{th}$ stator relative to the $\hat{x}$ axis [see Fig.~\ref{System}(a)], and $\kappa^{(1)}_{s,\,nj}$, $\kappa^{(2)}_{s,\,nj}$, and $\kappa^{(3)}_{s,\,nj}$ are positive stiffness coefficients. Again, in the special case where cylindrical rotors are rotating about their longitudinal axes, and stators are symmetric about the $\hat{x}$ axis, we find that $\kappa^{(2)}_{s,\,nj}$, and $\kappa^{(3)}_{s,\,nj}$ are constants and $\kappa^{(1)}_{s,\,nj}$ is zero. 

For application as magneto-mechanical transmitters \cite{Rhinithaa,UCLA} or wireless power transfer devices \cite{cook_dominiak_widmer_2009, 8234677}, an MMRA is typically actuated using a drive coil [Fig.~\ref{System}(b)]. Remarkably, we will show that this coil-based drive is responsible for additional coupling terms between the component oscillators, that modify the dynamics of the MMRA but have not been modeled previously. We can express the external torque, $\tau_{\textrm{ext},\,n}$, generated by the coil on the $n^\textrm{th}$ rotor with the expression \cite{Rhinithaa}:
\begin{equation}\label{tau_cn}
    {\tau}_{\textrm{ext},\,n} = \Gamma_n \text{cos}(\theta_{r,\,n}) I_c
\end{equation}
where $I_c$ is the instantaneous coil current and $\Gamma_n$ is the lumped coupling coefficient between the coil and the $n^\textrm{th}$ rotor that accounts for the magnetization and relative geometry of the rotor and coil. Each rotor simultaneously generates a back emf across the coil that can be expressed as \cite{Rhinithaa}
\begin{equation}\label{E_n}
    E_n = - \Gamma_n \text{cos}(\theta_{r,\,n}) \dot{\theta}_{r,\,n} ~.
\end{equation}
In a simplified case where the magnetic field generated by the coil at the $n^\textrm{th}$ rotor is uniformly distributed across the body of the rotor, the coupling coefficient can be expressed as 
\begin{equation}\label{Gamma_n}
    \Gamma_n = \frac{B_{r,\,n} V_n}{\mu_0} \frac{|{{B}_{c,\,n}}|}{|I_c|}
\end{equation}
where $B_{r,\,n}$ is the residual flux density of the $n^\textrm{th}$ rotor, $V_n$ is its volume, ${B}_{c,\,n}$ is the magnetic field generated by the coil [in $\hat{y}$ direction shown in Fig.~\ref{System}(b)] at the $n^\textrm{th}$ rotor due to the coil current $I_c$. For simplicity, we model the coil as an ideal inductor and can write the differential equation (in the undamped unforced form):
\begin{align}\label{eom_coil}
    L_c \frac{d I_c}{dt} = \sum\limits^{\substack{\textrm{all} \\ \textrm{rotors}}}_{i} E_i = - \sum\limits^{\substack{\textrm{all} \\ \textrm{rotors}}}_{i} \Gamma_i \text{cos}(\theta_{r,\,i}) \, \dot{\theta}_{r,\,i} ~.
\end{align}
Therefore, when $\dot{\theta}_{r,\,i} \neq 0 $,
\begin{align}\label{I_vs_theta}
    I_c = - \sum\limits^{\substack{\textrm{all} \\ \textrm{rotors}}}_{i} \frac{\Gamma_i}{L_c} \text{sin}(\theta_{r,\,i}) ~.
\end{align}
Substituting (\ref{I_vs_theta}) into (\ref{tau_cn}), we can get
\begin{equation}\label{tau_cn_complete}
    \tau_{\textrm{ext},\,n} = - \Gamma_n \text{cos}(\theta_{r,\,n}) \sum\limits^{\substack{\textrm{all} \\ \textrm{rotors}}}_{i}  \frac{\Gamma_i }{L_c} \text{sin}(\theta_{r,\,i}) ~.
\end{equation}
From (\ref{tau_cn_complete}), we can see that the presence of the coil adds indirect coupling terms between all rotors.

We can now explore the linear and nonlinear dynamics of the MMRA system. In the absence of any stators, all rotors will self-align in the same direction at equilibrium since this is the minimum energy configuration (in Fig.~\ref{System} this corresponds to the $\pm \hat{x}$ vector direction). Typically, in a magneto-mechanical transmitter \cite{Rhinithaa,UCLA}, all stators are set at $\theta_s = 0$ so that the torsional magnetic stiffness for the rotors is maximized. In the following analysis, we will accordingly fix the angular position of all stators to $\theta_s=0$. Other configurations can be studied using the same general method.

To study the dynamics of the MMRA, we expand (\ref{tau_r_general}), (\ref{tau_s_general}), and (\ref{tau_cn_complete}) using Taylor expansion:
\begin{align}
    \tau_{r,\,ni} 
        = & \nonumber - \left(
            \kappa^{(2)}_{r,\,ni} \theta_{r,\,i} 
                + \kappa^{(3)}_{r,\,ni} \theta_{r,\,n} 
                - \frac{1}{2}\kappa^{(2)}_{r,\,ni} \theta_{r,\,i} , \theta_{r,\,n}^2 \right. \notag \\
          &  \nonumber \quad \left. 
                - \frac{1}{2}\kappa^{(3)}_{r,\,ni} \theta_{r,\,i}^2 \theta_{r,\,n} 
                - \frac{1}{6}\kappa^{(2)}_{r,\,ni}  \theta_{r,\,i}^3 
                - \frac{1}{6} \kappa^{(3)}_{r,\,ni} \theta_{r,\,n}^3
            \right) \\
        & + H.O.T. \label{tau_r_taylor} \\
    \tau_{s,\,nj} 
        = & - \left(
            \kappa^{(3)}_{s,\,nj} \theta_{r,\,n} 
                - \frac{1}{6}\kappa^{(3)}_{s,\,nj}  \theta_{r,\,n}^3
            \right) + H.O.T. \label{tau_s_taylor} \\
    \tau_{\textrm{ext},\,n} 
        = &  
            - \Gamma_n \sum\limits_{i=1}^{N} \frac{\Gamma_i }{L_c} \left( \theta_{r,\,i} 
            - \frac{1}{2}\theta_{r,\,n}^2\theta_{r,\,i} 
            - \frac{1}{6}\theta_{r,\,i}^3 \right) + H.O.T. \label{tau_cn_taylor}
\end{align}
For analytical simplicity, we ignore the higher order terms ($H.O.T.$) that are greater than order 3. Later we will show that the resonance estimations are quite close to experimental reality even without these terms.

\subsection{Linear Dynamic Analysis}
\label{sec:LinearDynamic}

To study the linear dynamics of the MMRA, we can express (\ref{eom_rotor_general}) in the matrix form:
\begin{equation}\label{eom_matrix}
    J \ddot{\Theta}_r+ K\Theta_r = 0 ~.
\end{equation}
For an $N$-rotor system, the inertia matrix $J$ is a $N \times N$ diagonal matrix written as $J = \textrm{diag}(J_1,\,J_2,...,\,J_N)$ and $K$ is a $N \times N$ matrix that includes the torsional stiffness and the inter-rotor interactions. The vector $\Theta_r = [\theta_{r,\,1},\,\theta_{r,\,2},...,\,\theta_{r,\,N}]^T$ represents the  instantaneous angle of each rotor. The eigenfrequencies and corresponding mode shapes can be evaluated from the eigenvalues and eigenvectors of the matrix $J^{-1}K$.

We note that we are primarily interested in the in-phase mode of the MMRA in which all the rotors undergo synchronized motion. This is because the in-phase mode produces the largest net magnetic field and also couples best to the external magnetic environment (the other modes tend to be magnetically dark), which makes this mode particularly well suited for transmitter and receiver applications \cite{Rhinithaa,UCLA,cook_dominiak_widmer_2009}. However, for now, we will keep the analysis general.

For a linear analysis, we consider only the first order terms from (\ref{tau_r_taylor})-(\ref{tau_cn_taylor}) and expand the stiffness matrix $K$ into three contributions
\begin{equation}\label{K_matrix}
  K = K_\textrm{sus} + K_r + K_c ~.
\end{equation}
$K_\textrm{sus}$ is a diagonal matrix $K_\textrm{sus} = \textrm{diag}(\kappa_{\textrm{sus,}\,1},\,\kappa_{\textrm{sus,}\,2},..., \kappa_{\textrm{sus,}\,N})$ representing the stiffness component that comes from the suspension. 
$K_r$ is the stiffness component that comes from the other rotors and stators as described in (\ref{tau_r_taylor}) and (\ref{tau_s_taylor}), and can be expressed as
\begin{equation}\label{Kr_matrix}
  K_r(n,m) =
  \begin{cases}
    ~ \sum\limits^{\substack{\textrm{all} \\ \textrm{rotors}}}_{i} \kappa^{(3)}_{r,\,ni} + \sum\limits^{\substack{\textrm{all} \\ \textrm{stators}}}_{j} \kappa^{(3)}_{s,\,nj} & \text{, if $n = m$} \\
    ~ \kappa^{(2)}_{r,\,nm} & \text{, if $n \neq m$}
  \end{cases} ~.
\end{equation}
$K_\textrm{c}$ is the stiffness component that comes from the drive coil coupled to the system, as described in (\ref{tau_cn_taylor}), and can be expressed as
\begin{equation}\label{Kc_matrix}
  K_c(n,m) = \Gamma_n \frac{\Gamma_m}{L_c}  ~.
\end{equation}
From (\ref{Kc_matrix}), we can see that the coil will add extra stiffness that is quadratically proportional to the coil-rotor coupling factors $\Gamma$ and inversely proportional to the coil inductance $L_c$.

\subsection{Nonlinear Dynamic Analysis}
\label{sec:NonlinearDynamic}

In the previous analysis, we considered only the first order terms in the equation of motion. As oscillation amplitude increases, however, the effect of the higher order terms becomes significant and will modify the dynamics of the system. To understand the role of nonlinearities on the resonant frequencies and mode shapes of an MMRA, we can apply the first order harmonic balance method \cite{NNM}. Here, we consider $\theta_{r,\,n}$ as a single harmonic function
\begin{align}\label{harmonic_component}
    \theta_{r,\,n} = \alpha_n \cos(\omega t)
\end{align}
where $\alpha_n$ is the oscillation amplitude of the $n^\textrm{th}$ rotor and $\omega$ is the angular frequency. Using the identity $\cos^3(\omega t) = \frac{3}{4}\cos(\omega t) + \frac{1}{4}\cos(3 \omega t)$, we extract coefficients of only the $\cos(\omega t)$ terms in (\ref{tau_r_taylor})-(\ref{tau_cn_taylor}) by

\begin{align}
    \nonumber \tau_{r,\,ni,\, \omega} = & -\left(\kappa^{(2)}_{r,\,ni} \, \alpha_{i} + \kappa^{(3)}_{r,\,ni} \, \alpha_{n} -\frac{3}{8}\kappa^{(2)}_{r,\,ni} \, \alpha_i \alpha_n^2 \right.\\
    & \left. - \frac{3}{8}\kappa^{(3)}_{r,\,ni} \, \alpha_i^2 \alpha_n - \frac{1}{8}\kappa^{(2)}_{r,\,ni} \, \alpha_i^3 - \frac{1}{8} \kappa^{(3)}_{r,\,ni} \, \alpha_n^3 \right) \label{tau_ri_n_taylor_omega} \\
    \tau_{s,\,nj,\, \omega} = & -\left(\kappa^{(3)}_{s,\,nj} \, \alpha_{n} - \frac{1}{8}\kappa^{(3)}_{s,\,nj} \, \alpha_n^3 \right) \label{tau_si_n_taylor_omega} \\
    \tau_{c,\,n,\,\omega} = & - \Gamma_n \, \sum\limits^{\substack{\textrm{all} \\ \textrm{rotors}}}_{i} \frac{\Gamma_i }{L_c}\left(\alpha_i - \frac{3}{8}\alpha_n^2\alpha_i - \frac{1}{8}\alpha_i^3\right)  ~. \label{tau_cn_taylor_omega}
\end{align}
We consider here only the in-phase mode, in which all the rotors have synchronized motion with $\alpha_n > 0$. In this case, we find from ({\ref{tau_ri_n_taylor_omega}})-({\ref{tau_cn_taylor_omega}}) that the magnitude of the restoring torques provided by rotors, stators, and coil will decrease with the oscillation amplitude of rotors. This implies an amplitude-frequency softening characteristic for in-phase oscillations. For the $n^\textrm{th}$ rotor, balancing all coefficients of the cos($\omega t$) terms in the equation of motion, we obtain
\begin{align}\label{eom_omega}
     J_n\omega^2\alpha_n  = \kappa_{\textrm{sus},\,n}\alpha_n - \tau_{c,\,n,\,\omega} - \sum\limits^{\substack{\textrm{all} \\ \textrm{rotors}}}_{\substack{i \\ i \neq n}} \tau_{r,\,ni,\,\omega} - \sum\limits^{\substack{\textrm{all} \\ \textrm{stators}}}_{j} \tau_{s,\,nj,\,\omega} ~.
\end{align}
In principle, we can solve (\ref{eom_omega}) analytically for all rotors together to obtain $\alpha_n$ as a function of $\omega$. In practice, however, it is more convenient to find a numerical solution. Later in this manuscript, we will use (\ref{eom_omega}) to predict the amplitude-frequency response curves of the MMRA system.

\section{Eddy Current Losses Intrinsic to an MMRA}
\label{sec:Eddy}

In the above analysis, we are able to ignore losses associated with damping mechanisms as they do not, to leading order, modify the resonances or eigenmodes of the MMRA. However, these systems do have multiple sources of damping, which include various mechanical losses as well as eddy current damping. Typically, the dominant damping mechanism is associated with the mechanical suspensions for the rotors \cite{Rhinithaa,UCLA} and has been extensively discussed elsewhere \cite{Sarangi2004,JACOBS2014335,Nicholas1979,Armentrout1993,Kanj2022}. Eddy current damping originating from the time-varying magnetic fields and the moving conducting ferromagnetic materials (N52 neodymium alloy) also appears in MMRAs, but is relatively unexplored. In particular, for MMRAs that have compactness as a design goal, the eddy current effect becomes quite significant. Here we adopt an equivalent circuit approach \cite{Eddy_current_model1,Eddy_current_model2} to study the nature of this eddy current damping, and additionally show that the eddy current losses can be well-estimated with a simplified model, without reliance on complex finite element simulations.

\begin{figure}[t!] 
    \centering
  {\includegraphics[width=1\linewidth]{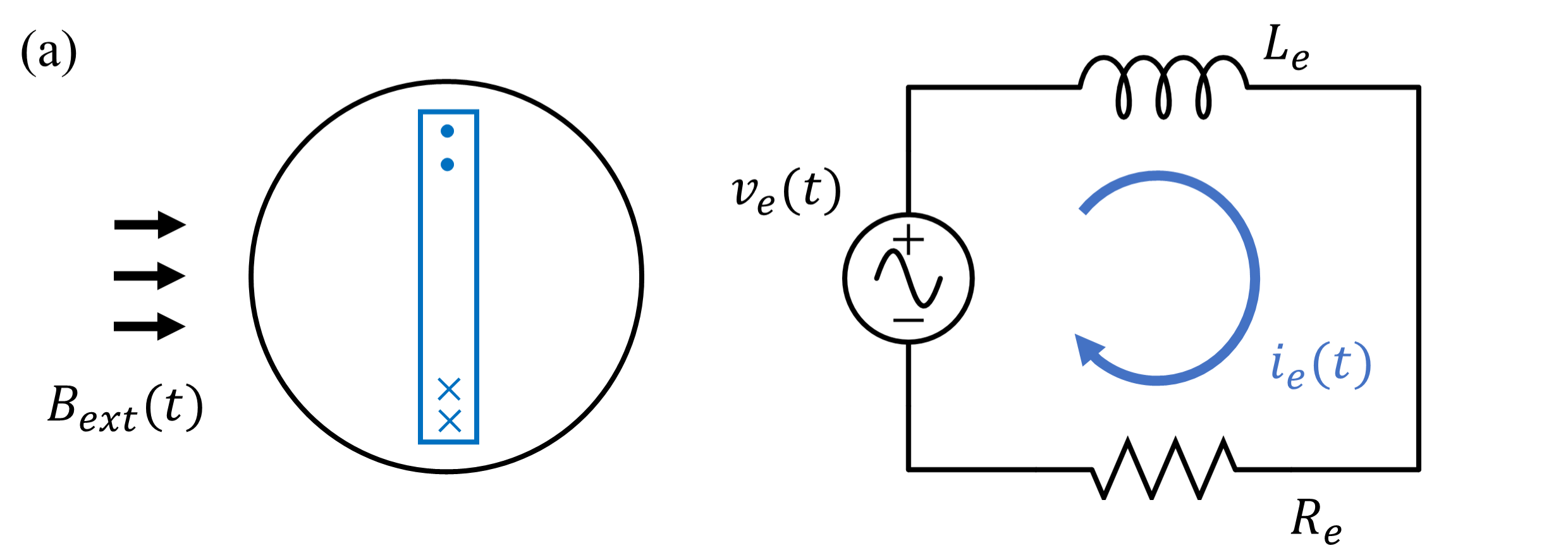}}
\\
 {\includegraphics[width=1\linewidth]{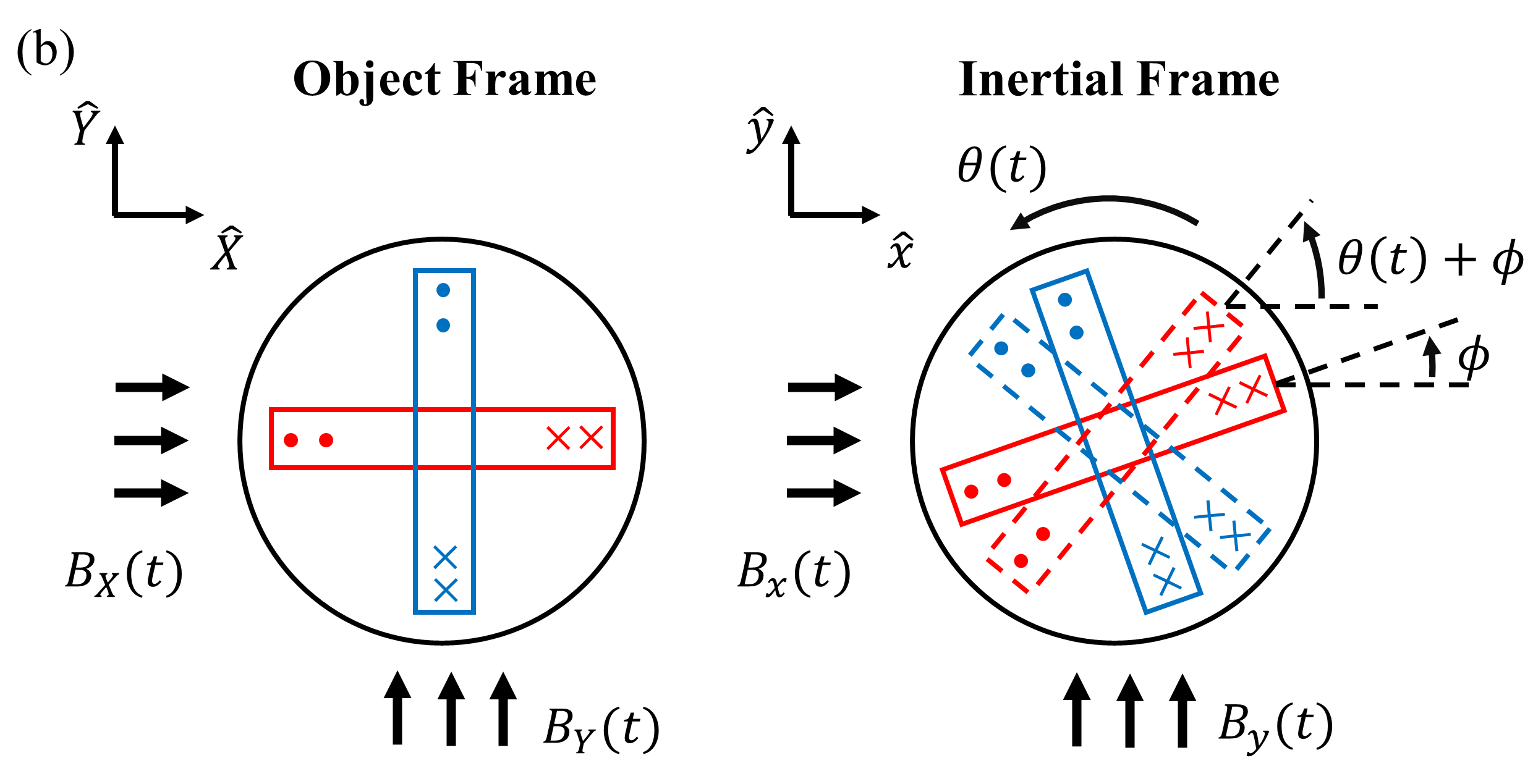}}
\caption{\textbf{(a)} Schematic of a conducting object experiencing a time-varying magnetic field (left), $B_\textrm{ext}(t)$, and equivalent circuit for the eddy current (right). The blue solid line encircles the eddy current loop in the conductor. The crosses and dots represent the positive flow direction of the eddy current. $v_e(t)$, $i_e(t)$, $L_e$, and $R_e$ represent the effective time-varying induced voltage, time-varying induced current, inductance, and resistance for the circuit. \textbf{(b)} Schematic of the eddy current in the object frame (left) and inertial frame (right). The blue and red solid lines encircle the eddy current loops with normal vectors pointing in the $\hat{X}$ and $\hat{Y}$ direction, respectively. $\phi$ is the initial at-rest angle between the object frame and the inertial frame. The blue and red dash lines encircle the same loops in the inertial frame after the object rotating by $\theta(t)$.
}
  \label{eddy_current_model} 
\end{figure}

Since time-varying magnetic fields produce eddy current loops within conductors, a coil-based circuit model [Fig.~\ref{eddy_current_model}(a)] is routinely adopted, with the induced current being modeled as the result of an effective potential across the coil. This effective induced voltage in the circuit can be expressed as
\begin{align}
    v_{e} 
        =  - \frac{d{(B_\textrm{ext}{A}_{e})}}{dt}  \label{v_e}
\end{align}
where $B_\textrm{ext}$ is the external magnetic flux density and $A_e$ is the effective area of the equivalent circuit. The effective area depends on the geometry of the conductor in the direction along the magnetic field. If the time-variation is harmonic, the instantaneous eddy current loss can be generally expressed as:
\begin{align}
    P_{e}
        = \frac{v_{e}^2R_{e}}{R_{e}^2 + (\omega_b L_{e})^2} \label{peddy}
\end{align}
where $\omega_b$ is the angular frequency of the external magnetic flux density, $R_e$ and $L_e$ are the effective resistance and inductance of the coil model, respectively. These effective parameters also depend on the geometry. In the context of MMRAs, we are typically only concerned with operation at the ultra-low frequency (ULF, 0.3 to 3 kHz) range \cite{Rhinithaa, UCLA} or below. Therefore, we may employ the quasi-static approximation such that $\omega_b^2 L_e^2 \ll R_e^2$, and the eddy current loss can be simplified as purely resistive:
\begin{align}
    P_{e} = \frac{v_{e}^2}{R_{e}} ~. \label{peddy_simplify}
\end{align}
As we will show later, this quasi-static assumption can provide a fairly good estimation of the eddy current loss in MMRAs operating at ULF. 

A conducting object, e.g., a rotor or stator, within the MMRA experiences a time-varying magnetic flux density that can be decomposed into $B_x\hat{x}$ and $B_y\hat{y}$ components in the inertial reference frame. We can therefore consider the conducting object as two coils with perpendicular normal vectors. During operation, some conducting objects, e.g., rotors, will have time-varying angular positions resulting in time-varying effective area $A_e$ which may complicate the analysis. We can reduce this complexity by defining a new coordinate system $\hat{X}$ and $\hat{Y}$ as the co-moving frame of reference of the object, as shown in Fig.~\ref{eddy_current_model}(b), such that the effective loop areas do not vary in time. Since we can choose the object frame in the direction where the effective parameters of current loops are convenient, we allow for an initial at-rest angle $\phi$ between the object frame and the inertial frame. In this object frame, the magnetic flux density permeating the conductor $B_X\hat{X}+B_Y\hat{Y}$ can be expressed as
\begin{align}
    B_{X} 
        = & B_x\,\textrm{cos}(\theta + \phi) + B_y\,\textrm{sin}(\theta + \phi)\label{B_X}\\
    B_{Y} 
        = & B_y\,\textrm{cos}(\theta + \phi) - B_x\,\textrm{sin}(\theta + \phi) \label{B_Y}
\end{align}
where $\theta$ is the instantaneous rotation angle of the object in the inertial frame. The instantaneous eddy current loss in each circuit (in the $\hat{X}$ and $\hat{Y}$ direction) can now be expressed as
\begin{align}
    P_{e,\,X/Y} = & \frac{v_{e,\,X/Y}^2}{R_{e,\,X/Y}} = \frac{A_{e,\,X/Y}^2}{R_{e,\,X/Y}} \dot{B}_{X/Y}^2 \label{peddy_XY} ~.
\end{align}

As stated previously, the coefficients $\frac{A_{e, X/Y}^2}{R_{e, X/Y}}$ are effective properties for the current loop model and therefore depend on the the geometry along the corresponding directions. For a cylindrical conducting object, the eddy current loss coefficient along its radial direction can be analytically derived from \cite{Aubert2012} and written as (see details in Appendix \ref{sec:Eddy_current_loss_coeff})
\begin{align}\label{Effective_A_rotor}
    \frac{A_e^2}{R_e} = \frac{128}{\pi^3} \sigma a^2 b^3 \sum_{i} \frac{1}{(2i+1)^4} \frac{J_2(ma)}{J_0(ma) + J_2(ma)}
\end{align} 
where $\sigma$ is the conductivity, $a$ is the radius, $b$ is half of the length, $J_0$ and $J_2$ are the Bessel functions of the first kind, $i = 0, 1, ...$, and $m = \frac{(2i+1)\pi}{2b}$. For a cuboidal conducting object, the eddy current loss coefficient along the corresponding surface normal vector can also be analytically derived from {\cite{767162}} and written as (see details in the Appendix \ref{sec:Eddy_current_loss_coeff})
\begin{align}\label{Effective_A_stator}
    \frac{A_e^2}{R_e} = \frac{\sigma d_t^2}{12} V_\textrm{cub}
\end{align} 
where {$d_t$} is the thickness along the corresponding direction and {$V_\textrm{cub}$} is the volume of the cuboidal conducting object. For other arbitrary geometries of the rotors, it is more convenient to find the eddy current loss coefficient using a numerical simulation or from an experiment. In particular, in simulation (or experiment), we can apply a uniform time-varying magnetic flux density $B(t) = B_o\,\textrm{cos}(\omega_b t)$, where $B_o$ is a constant, passing through the conductor along the direction in which we are interested in finding the eddy current loss coefficient. According to (\ref{v_e}) and (\ref{peddy_simplify}), the time-varying eddy current loss would be of the form $P_e(t) = P_o \, \textrm{sin}^2(\omega_b t) $, where $P_o$ is a constant under the quasi-static approximation and can be found in simulation (or experiment). We can then obtain the eddy current loss coefficient in that direction by computing
\begin{align}\label{A_eff_num}
    \frac{A_{e}^2}{R_{e}} ={\frac{P_o}{\omega_b^2B_o^2}} ~.
\end{align}

We note, for the sake of improved accuracy, that the magnetic flux density permeating each conducting object within the MMRA is not spatially uniform but it can be analytically / numerically estimated with some modest effort [see (\ref{Appendix_dB_xyz}) in Appendix \ref{sec:Magnetic_Torque}]. We can, however, make a simplification here by averaging the magnetic flux density along the desired vector direction across the entire object and then using it as an equivalent uniform local value. This average magnetic flux density on the $n^\textrm{th}$ rotor in $\hat{x}$ and $\hat{y}$ direction can be generally expressed as (see Appendix \ref{sec:Magnetic_Torque}):
\begin{align}
    B_{x,\,n}
        = \nonumber & \sum\limits^{\substack{\textrm{all} \\ \textrm{rotors}}} _{\substack{i \\ i \neq n}} \,  \left[b^{(1)}_{r,\,ni}  \textrm{cos}(\theta_i) + b^{(2)}_{r,\,ni}\,\textrm{sin}(\theta_i)\right] \\ & +  \sum\limits^{\substack{\textrm{all} \\ \textrm{stators}}} _{j} \,  \left[b^{(1)}_{s,\,nj} \, \textrm{cos}(\theta_j) + b^{(2)}_{s,\,nj}\,\textrm{sin}(\theta_j)\right] \label{Bx_n}\\
    B_{y,\,n}
        = \nonumber & \sum\limits^{\substack{\textrm{all} \\ \textrm{rotors}}} _{\substack{i \\ i \neq n}} \,  \left[b^{(2)}_{r,\,ni} \, \textrm{cos}(\theta_i) + b^{(3)}_{r,\,ni}\,\textrm{sin}(\theta_i)\right] \\ & + \sum\limits^{\substack{\textrm{all} \\ \textrm{stators}}} _{j} \,  \left[b^{(2)}_{s,\,nj} \, \textrm{cos}(\theta_j) + b^{(3)}_{s,\,nj}\,\textrm{sin}(\theta_j)\right] \label{By_n}
\end{align}
where $b^{(1)}$, $b^{(2)}$, and $b^{(3)}$ are coefficients that are functions of angles and further depend on the geometry of the rotors and stators. Again, in the special case where cylindrical rotors are rotating about their longitudinal axes, and stators are symmetric about the $\hat{x}$ axis (with stator magnetization $\theta_s = 0$), we find that $b^{(1)}$ and $b^{(3)}$ become constants, and $b^{(2)}$ becomes zero [see (\ref{Appendix_b1_to_b3}) and (\ref{Appendix_C1_to_C4}) in Appendix \ref{sec:Magnetic_Torque}]. Therefore, we can rewrite a simplified form of (\ref{Bx_n}) and (\ref{By_n}) as
\begin{align}
    B_{x,\,n}
        = & \sum\limits^{\substack{\textrm{all} \\ \textrm{rotors}}} _{\substack{i \\ i \neq n}} \,  b^{(1)}_{r,\,ni} \, \textrm{cos}(\theta_i)  +  \sum\limits^{\substack{\textrm{all} \\ \textrm{stators}}} _{j} \,  b^{(1)}_{s,\,nj}   \label{Bx_n_simplify}\\
    B_{y,\,n}
        = & \sum\limits^{\substack{\textrm{all} \\ \textrm{rotors}}} _{\substack{i \\ i \neq n}} \,  b^{(3)}_{r,\,ni}\,\textrm{sin}(\theta_i) \label{By_n_simplify} ~.
\end{align}
The average magnetic flux density in the stators can be analytically found following the same general method.  

\section{Results and Discussion}
\label{sec:Discussion}
In this section, we present experimental and numerical results from two different MMRA prototypes: a single-rotor MMRA and a multi-rotor MMRA.

\subsection{Prototypes and Measurement Setup}
\begin{figure*}
    \centering
  \includegraphics[width=1\linewidth]{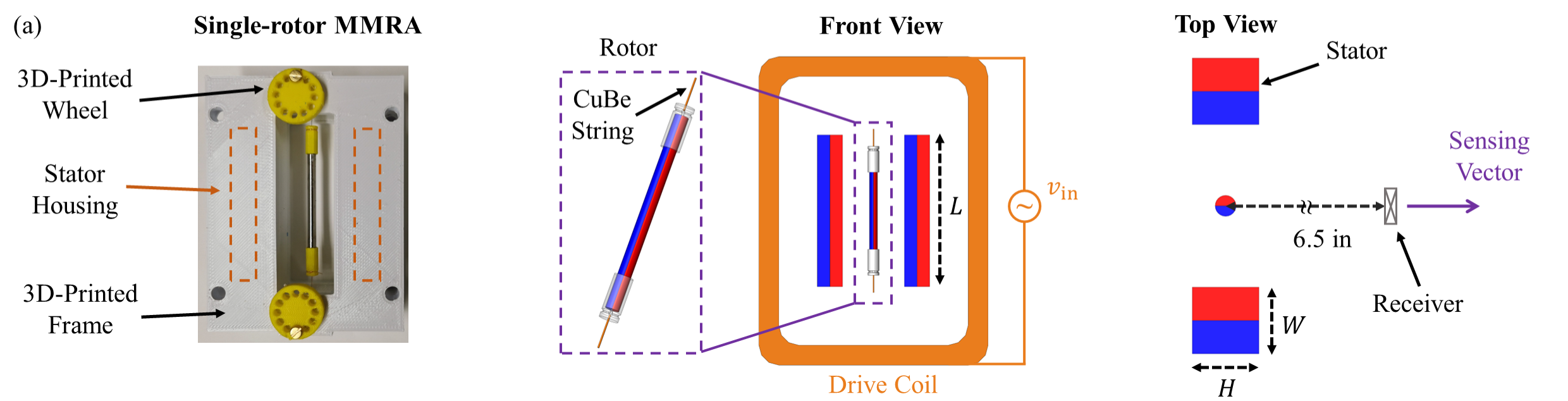}
\\
 \includegraphics[width=1\linewidth]{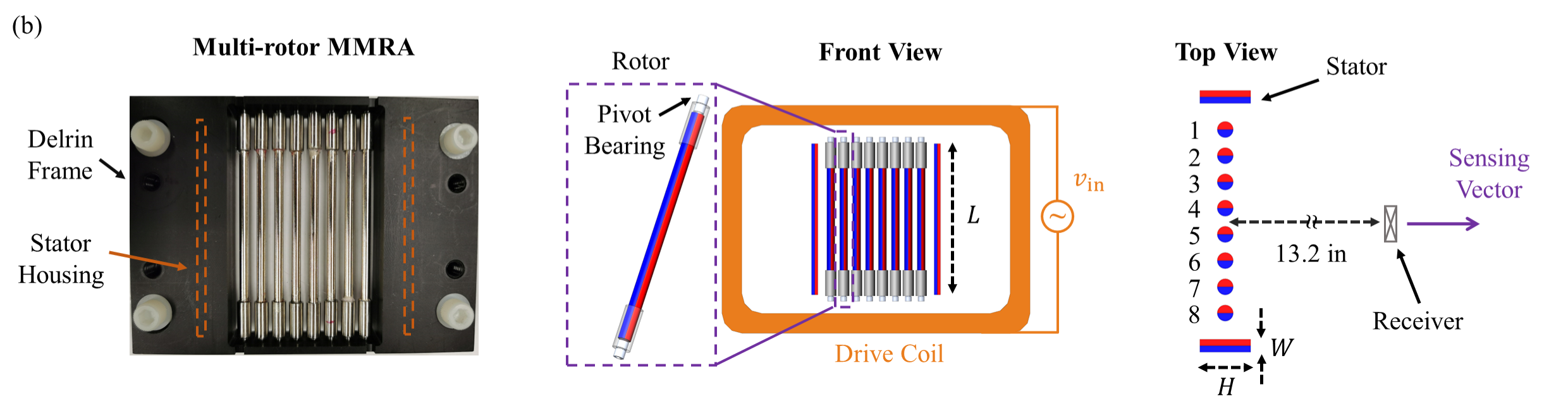}
\caption{
    Photographs and schematics of {\textbf{(a)}} a single-rotor MMRA and {\textbf{(b)}} a multi-rotor MMRA coupled to a drive coil. In the multi-rotor MMRA, rotors are equally spaced. In both figures, the length, width, and height of stators are denoted by $L$, $W$, and $H$, respectively. In (b), rotors in the multi-rotor MMRA are labeled from 1 to 8.}
  \label{Devices} 
\end{figure*}

Fig.~\ref{Devices} presents our experimental prototypes. For both prototypes, we use N52 grade neodymium magnets (K\&J Magnetics, Inc.) for the rotors and stators. The single-rotor MMRA shown in Fig.~\ref{Devices}(a) consists of a 3D-printed frame with a cylindrical rotor magnet suspended using metal strings (0.25 mm diameter Copper Beryllium) under tension. The multi-rotor MMRA [Fig.~\ref{Devices}(b)] consists of a Delrin frame with cylindrical rotor magnets suspended using flexure bearings (C-Flex Bearing Co., Inc. Single End Bearing A-10). Stators for both prototypes are secured inside the respective frames. In Appendix \ref{sec:Values_of_Key_Parameters}, we provide values of all key parameters for both MMRA prototypes. We use a 100-turn coil (95 mm $\times$ 135 mm inner dimensions) made of AWG 18 enameled copper wire with sinusoidal voltage input to actuate the MMRAs during experiments. A flux-gate magnetometer (Texas Instruments DRV425EVM) is used as the receiver to measure the total magnetic field generated by the MMRAs and the coil. Using the same receiver configuration, we calibrate the coil independently of the MMRAs, and are able to subtract the phasor field generated by the coil from the total phasor field measured at the receiver to get the phasor field generated by the MMRAs. In the experiment, the average power consumption in the MMRA is calculated by measuring the total power consumption using voltage across the coil in conjunction with the coil current and subtracting the coil power consumption from the total power consumption {\cite{Rhinithaa}}.

\subsection{Frequency Response of the MMRA Prototypes}
\label{sec:FrequencyResponse}
\begin{figure} [t]
    \centering
    {\includegraphics[width=\linewidth]{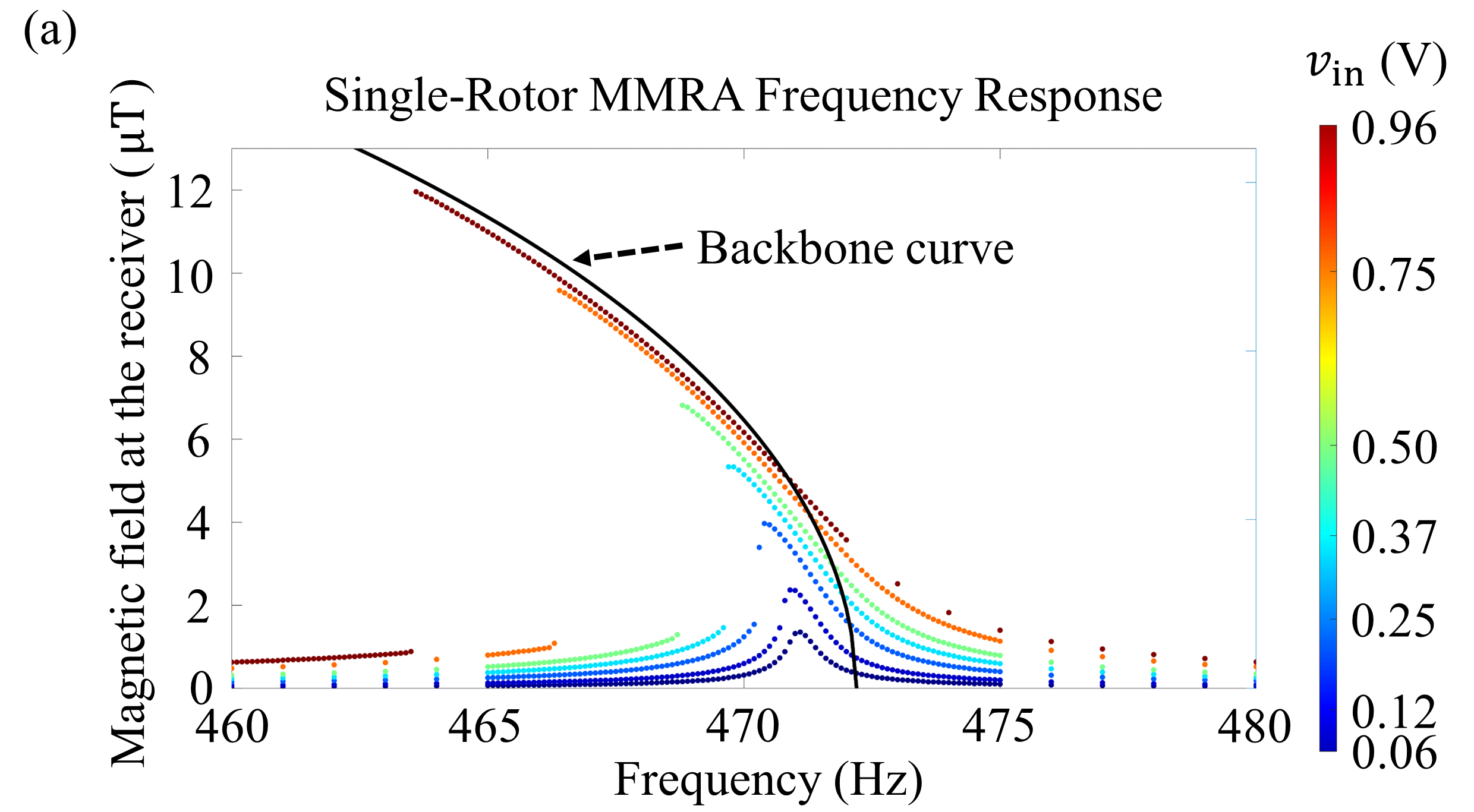}}
    \\
    {\includegraphics[width=\linewidth]{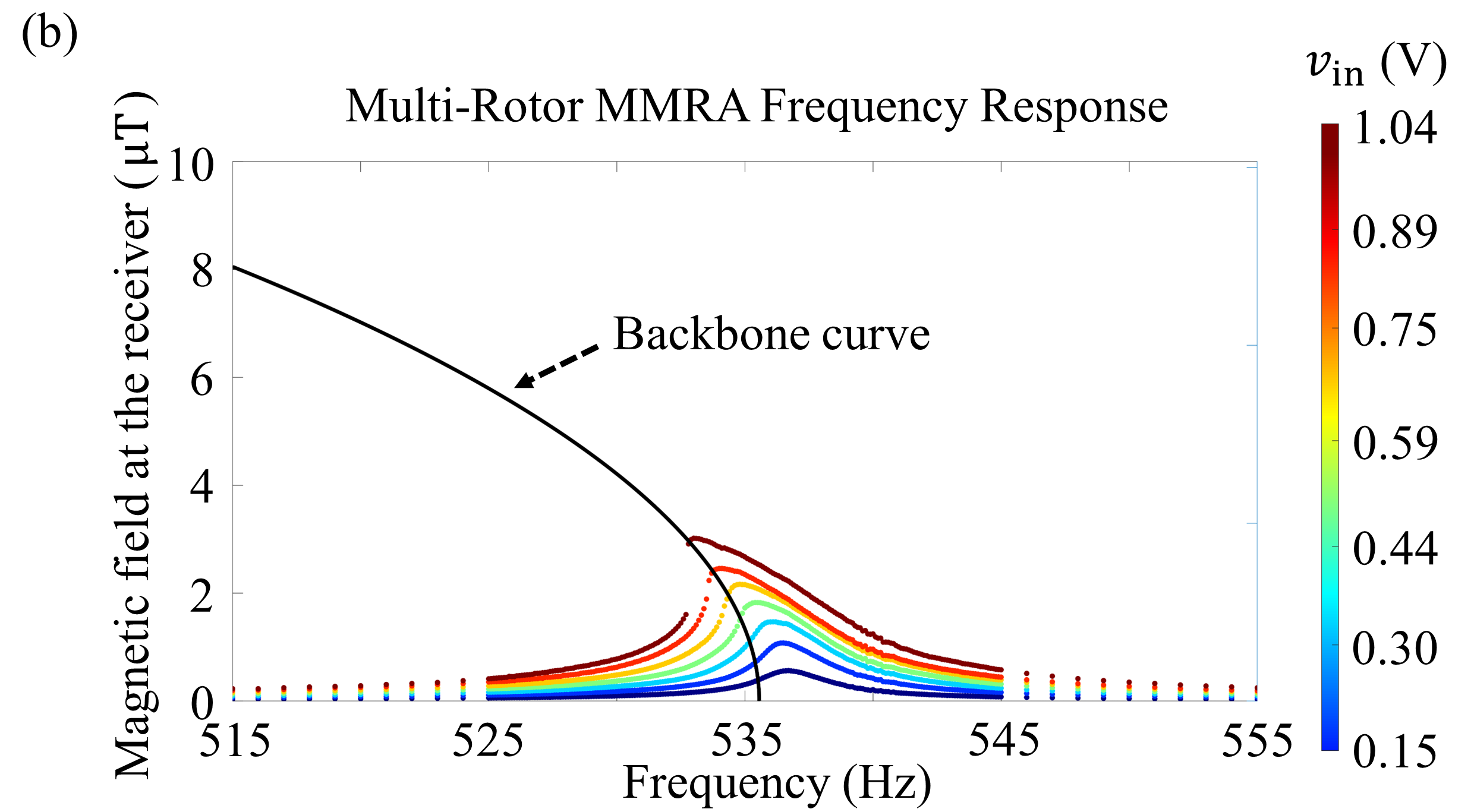}}
    \\
    \caption{
        Measured frequency responses of {\textbf{(a)}} the single-rotor MMRA and {\textbf{(b)}} the multi-rotor MMRA during downward sweep of the drive frequency at various drive voltages, whose rms value is denoted by {$v_\textrm{in}$}. The Y-axis represents the rms value of magnetic field measured at the receiver.
    }
  \label{Dyanmic_results} 
\end{figure}

Fig.~\ref{Dyanmic_results} shows measured frequency response curves for the in-phase mode of the MMRA prototypes at various drive voltages. Since we expect softening nonlinear behavior, in which resonant frequencies decrease with increasing amplitude, we perform sweeps with decreasing drive frequency to fully capture the softening characteristic.

\begin{table}[!t]
	\renewcommand{\arraystretch}{1.3}
	\caption{Comparison between linear resonant frequencies}
	\label{Natural_frequency_results}
	\resizebox{\columnwidth}{!}{
		\begin{tabular}{c c c c}
			\hline\hline \\[-3mm]
			MMA        &Frequency from   &Frequency from      &Frequency from\\
prototype        &the experiment       &the analytical model   &the dipole model  \\[1.6ex] \hline
Single-rotor            & 471.2 Hz        & 472.2 Hz           & 860.3 Hz \\ 
Multi-rotor$^{\,\ast}$            & 536.8 Hz        & 535.6 Hz           & 1454.4 Hz   \\ 
			\hline\hline
		\end{tabular}
	}
	
	\begin{tabular}{l}
         $^{\,\ast}$ Frequency corresponds to the in-phase mode.
    \end{tabular}

\end{table}

Experimentally, we observe the softening characteristic in both MMRAs as predicted. Using (\ref{eom_omega}) (with the values of parameters from Table~\ref{Dimension_and_parameter} in Appendix \ref{sec:Values_of_Key_Parameters}), we can analytically estimate the oscillation amplitude of each rotor as a function of the resonant frequency. With the support of (\ref{Appendix_Bx})-(\ref{Appendix_b1_to_b3}) in Appendix \ref{sec:Magnetic_Torque}, we can then predict the amplitude of the net magnetic field generated at the receiver as a function of the resonant frequency including the nonlinear softening effect, which gives us the ``backbone curves'' shown in Fig.~\ref{Dyanmic_results}. We find that there is very good agreement between the analytical model and the experimental data, except for a slight frequency offset. In Table~\ref{Natural_frequency_results}, we summarize the measured resonant frequencies at low amplitude (linear regime), and compare them to the analytical linear resonant frequencies estimated using our improved analytical model presented in \S\ref{sec:LinearDynamic} and the less accurate dipole model described in (\ref{tau_dipole}). Notably, the dipole model greatly overestimates the linear resonant frequency of MMRAs since it fails to consider the shape and near field effects of magnetic oscillators. The resonant frequencies of these MMRA prototypes found using our improved analytical model, on the other hand, are in very close agreement with those measured in experiments. Therefore, it is clear that this analytical model tremendously improves our predictive capabilities for MMRAs, especially in the regime where the spacing between adjacent magnetic oscillators is comparable to their cross-section. 

For the multi-rotor MMRA, we can additionally predict the oscillation mode shape of the in-phase mode as a function of the resonant frequency in the nonlinear regime (see Appendix \ref{sec:Nonlinear_Mode_Shape}). Here, we see that the mode shape distributes a little more evenly across the resonators as the oscillation amplitude increases, which is advantageous for power considerations \cite{Rhinithaa}.

\subsection{Eddy Current Loss Estimation}

Since it is not practical to experimentally distinguish the eddy current loss from other mechanical losses, we must rely on simulations to estimate the fraction of the measured loss that is contributed by the eddy currents within the MMRA. Since this is an involved exercise, we also aim to now compare the predictions of our simplified eddy current loss model (described in \S\ref{sec:Eddy}) to the estimation from finite element simulations, thereby confirming the utility of the simplified model.

\begin{figure} 
    \centering
  {\includegraphics[width=\linewidth]{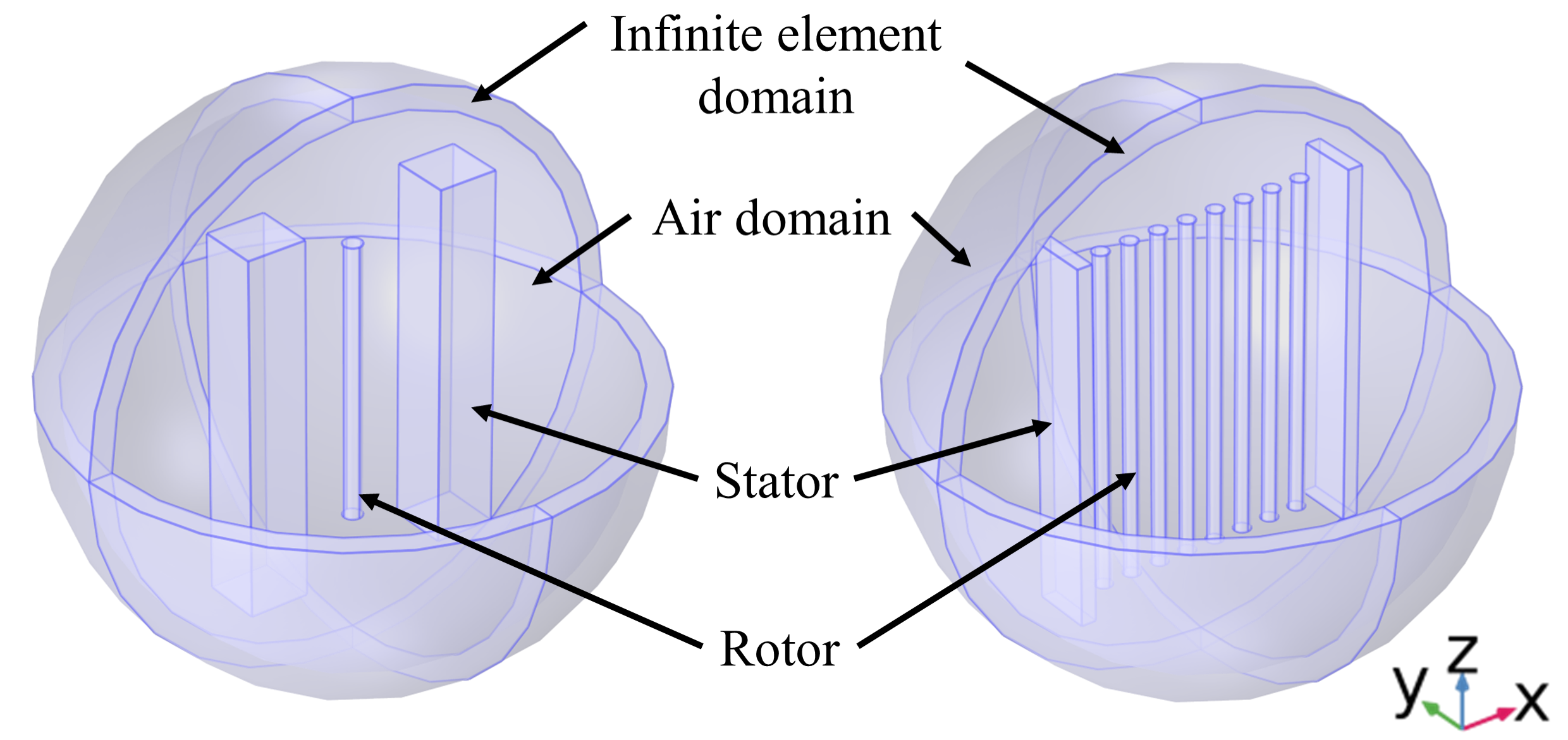}}
 
    \caption{3D models developed in COMSOL to perform numerical simulations of the single-rotor MMRA (left) and the multi-rotor MMRA (right).
    }
  \label{Eddy_current_models} 
\end{figure}

We perform numerical simulations of the single-rotor MMRA and multi-rotor MMRA using COMSOL Multiphysics. The overall 3D models created in COMSOL are illustrated in Fig.~{\ref{Eddy_current_models}}, both of which comprise cylindrical conductors representing the rotors, rectangular cuboids representing the stators, and a spherical domain encompassing all the components representing the air. All dimensions and material properties used in these models are shown in Appendix \ref{sec:Values_of_Key_Parameters} Table~\ref{Dimension_and_parameter}. The sizes of the air domains are set to optimize the run-time of the simulation and an infinite element domain is applied to simulate an infinitely extended air domain. In both simulations, rotors are modeled as oscillating relative to the reference frame so that we can use a ready-made physics interface for rotating machinery to model the oscillation of rotors.

\begin{table}[!t]
	\renewcommand{\arraystretch}{1.3}
	\caption{Average eddy current loss in the single-rotor MMRA}
	\centering
	\label{single_rotor_eddy_current_results}
	\resizebox{\columnwidth}{!}{
		\begin{tabular}{c c c c}
			\hline\hline \\[-3mm]
			Single-rotor                    & Oscillation                     & Analytically estimated                & Numerically estimated\\
        MMRA                            & amplitude                       & eddy current loss                     & eddy current loss\\[1.6ex] \hline
          Stator                        & N.A.                            & 1.64 {mW}                                & 1.49 {mW}    \\
          Rotor                         & 30 {deg}                          & 11.54 {mW}                               & 11.73 {mW}   \\
			\hline\hline
		\end{tabular}
	}
\end{table}

\begin{table}[!t]
	\renewcommand{\arraystretch}{1.3}
	\caption{Average eddy current loss in the Multi-rotor MMRA}
	\centering
	\label{multi_rotor_eddy_current_results}
	\resizebox{\columnwidth}{!}{
		\begin{tabular}{c c c c}
			\hline\hline \\[-3mm]
			Multi-rotor                     & Oscillation                     & Analytically estimated               & Numerically estimated\\
        MMRA                            & amplitude                       & eddy current loss                    & eddy current loss\\[1.6ex] \hline
                    Stator                        & N.A.                            & 0.38 {mW}                               & 0.28 {mW}   \\
          Rotor 1 $\&$ 8                & 4.8 {deg}                            & 0.84 {mW}                               & 0.81 {mW}   \\
          Rotor 2 $\&$ 7                & 7.6 {deg}                            & 2.13 {mW}                               & 2.00 {mW}   \\
          Rotor 3 $\&$ 6                & 9.0 {deg}                            & 3.06 {mW}                               & 2.86 {mW}   \\
          Rotor 4 $\&$ 5                & 9.7 {deg}                             & 3.51 {mW}                               & 3.28 {mW}   \\
			\hline\hline
		\end{tabular}
	}
\end{table}

Since we are trying to estimate the fraction of the measured average power consumption that is contributed by the eddy currents within the MMRAs, in the simulation, we apply prescribed sinusoidal angular motions on the rotors in accordance with the experimental results. For the single-rotor MMRA, we choose the experimental result at the experimental resonant frequency at 463.1 Hz [Fig.~{\ref{Dyanmic_results}(a)}]. The experimental oscillation amplitude of the rotor (provided in Table~{\ref{single_rotor_eddy_current_results}}) is estimated from the magnetic field experimentally measured at the receiver with the support of (\ref{Appendix_Bx})-(\ref{Appendix_b1_to_b3}) in Appendix \ref{sec:Magnetic_Torque}. For the multi-rotor MMRA, we choose the experimental result at the experimental resonant frequency at 532.8 Hz [Fig.~{\ref{Dyanmic_results}(b)}]. The oscillation mode shape for this multi-rotor case is similarly analytically estimated (as described in {\S\ref{sec:NonlinearDynamic}} and Appendix \ref{sec:Nonlinear_Mode_Shape}), which can be used in conjunction with the magnetic field measured at the receiver, to estimate the oscillation amplitude for each rotor (see Table~{\ref{multi_rotor_eddy_current_results}}). We simulate both MMRAs for one period of oscillation to calculate the average eddy current loss in each conductor in the MMRAs.

As described in {\S \ref{sec:Eddy}}, the simplified eddy current loss model relies on the eddy current loss coefficient {$A_{e}^2/R_{e}$} associated with each conducting object. While ({\ref{Effective_A_rotor}}) and ({\ref{Effective_A_stator}}) model this coefficient explicitly for a cylindrical conductor along the radial direction and a cuboidal conductor along the corresponding surface normal vector, which are the cases for the rotors and stators in our prototypes, respectively, we can alternatively use the method described alongside (\ref{A_eff_num}) to numerically estimate this eddy current loss coefficient for an arbitrary geometry conductor. In Table~\ref{cylindrical_eddy_current_coefficient} and Table~\ref{cuboidal_current_coefficient}, we calculate the eddy current loss coefficients for the cylindrical rotors and cuboidal stators, as used in our MMRA prototypes, using both the analytical methods as well as the numerical approach. For both rotors and stators used in our prototypes, we find that the eddy current loss coefficient estimations from the analytical methods are very well matched to the estimations from the numerical method. As discussed in (\ref{Effective_A_rotor}), the eddy current loss coefficient for a cylindrical rotor is quadratically proportional to its radius and cubically proportional to its length. Similarly, as discussed in ({\ref{Effective_A_stator}}), the eddy current loss coefficient for a cuboidal stator is quadratically proportional to its thickness in the corresponding direction and linearly proportional to its volume. Therefore, to reduce eddy current loss in MMRAs, the best options are to either use thinner and shorter, or even laminated, rotor and stator magnets.

\begin{table}[!t]
	\renewcommand{\arraystretch}{1.3}
	\caption{Comparison between the eddy current loss coefficients for cylindrical rotors}
	\centering
	\label{cylindrical_eddy_current_coefficient}
	\resizebox{\columnwidth}{!}{
		\begin{tabular}{c c c}
			\hline\hline \\[-3mm]
        Parameter                                                       & Single-rotor                                           & Multi-rotor\\
        ~                                                               & MMRA                                                   & MMRA\\
\hline
          Operating frequency              & 463.1 {Hz}                                               & 532.8 {Hz}     \\
          Radius                                                  & 2.00 {mm}                                                & 1.75 {mm}    \\
          Length                                                  & 2.3125 {in}                                              & 2.8125 {in}    \\
          Conductivity                                            & 0.625 {MS/m}                                             & 0.625 {MS/m}     \\
          Analytical {$\frac{A_{e}^2}{R_{e}}$}                      & 0.46 {$\mu \textrm{W}\textrm{s}^2/\textrm{rad}^2\textrm{T}^2$}     & 0.33 {$\mu \textrm{W}\textrm{s}^2/\textrm{rad}^2\textrm{T}^2$}      \\
          Numerical {$\frac{A_{e}^2}{R_{e}}$}                       & 0.44 {$\mu \textrm{W}\textrm{s}^2/\textrm{rad}^2\textrm{T}^2$}     & 0.32 {$\mu \textrm{W}\textrm{s}^2/\textrm{rad}^2\textrm{T}^2$}      \\

			\hline\hline
		\end{tabular}
	}
\end{table}

\begin{table}[!t]
	\renewcommand{\arraystretch}{1.3}
	\caption{Comparison between the eddy current loss coefficients for cuboidal stators}
	\centering
	\label{cuboidal_current_coefficient}
	\resizebox{\columnwidth}{!}{
		\begin{tabular}{c c c}
			\hline\hline \\[-3mm]
        Parameter                                                       & Single-rotor                                           & Multi-rotor\\
        ~                                                               & MMRA                                                   & MMRA\\
\hline
        Operating frequency              & 463.1 {Hz}                                               & 532.8 {Hz}     \\

Length                                                & 3 {in}                                                   & 3 {in} \\
          Width                                                & 0.5 {in}                                                 & 0.125{in} \\
          Height                                                 & 0.5 {in}                                                 & 0.5 {in} \\
          Conductivity                                           & 0.625 {MS/m}                                             & 0.625 {MS/m}     \\
          Analytical {$\frac{A_{e}^2}{R_{e}}$} along width         & 103.2 {$\mu \textrm{W}\textrm{s}^2/\textrm{rad}^2\textrm{T}^2$}    & 1.61 {$\mu \textrm{W}\textrm{s}^2/\textrm{rad}^2\textrm{T}^2$}      \\
          Numerical {$\frac{A_{e}^2}{R_{e}}$} along width          & 92.7 {$\mu \textrm{W}\textrm{s}^2/\textrm{rad}^2\textrm{T}^2$}     & 1.64 {$\mu \textrm{W}\textrm{s}^2/\textrm{rad}^2\textrm{T}^2$}                         \\
          Analytical {$\frac{A_{e}^2}{R_{e}}$} along height        & 103.2 {$\mu \textrm{W}\textrm{s}^2/\textrm{rad}^2\textrm{T}^2$}    & 25.8 {$\mu \textrm{W}\textrm{s}^2/\textrm{rad}^2\textrm{T}^2$}      \\
          Numerical {$\frac{A_{e}^2}{R_{e}}$} along height         & 92.7 {$\mu \textrm{W}\textrm{s}^2/\textrm{rad}^2\textrm{T}^2$}     & 23.2 {$\mu \textrm{W}\textrm{s}^2/\textrm{rad}^2\textrm{T}^2$}       \\
			\hline\hline
		\end{tabular}
	}
\end{table}

Table \ref{single_rotor_eddy_current_results} and Table \ref{multi_rotor_eddy_current_results} compare the numerically and analytically estimated average eddy current loss in all conductors in both MMRAs. As shown in both tables, the eddy current loss estimations are in good agreement for each conductor. The slight discrepancies arise from the fact that in the simplified model we assume that the magnetic field generated by one magnet on another is uniform, which is less valid as the magnets become closer to each other. In both single-rotor and multi-rotor MMRAs, we find that most of the eddy current loss occurs in the rotors. Moreover, most of the eddy current loss occurs in the rotors that are closer to the center since they experience a larger magnetic field generated by the other magnets and also undergo oscillation with a larger amplitude.

In our experiments, we find that the average power consumption in the single-rotor MMRA at 463.1 Hz and multi-rotor MMRA at 532.8 Hz are 79.4 mW and 219.9 mW, respectively. Based on our modeling, we estimate that in the single-rotor MMRA the eddy current loss contributes about $18.5\%$ of the total, and in the multi-rotor MMRA the eddy current loss contributes about $8.9\%$ of the total. This leads to the conclusion that, in these two prototypes, most of the power loss occurs in the mechanical suspension. The eddy current loss, however, is inherently present in all MMRA systems and sets the lower bound on lowest possible power consumption for such systems.

\section{Conclusion}

In this work we have developed a new analytical dynamical model for generalized torsional magneto-mechanical resonator arrays. We demonstrate that this analytical approach tremendously improves our frequency and oscillation mode prediction capability for MMRAs in the linear regime, as well as in the nonlinear regime, with a direct comparison to experimental results. We additionally develop a new simplified model that can be used to estimate the eddy current loss in MMRAs, without reliance on complicated finite element simulations. This model reveals that the eddy current loss in each conductor in the experimental MMRA prototypes estimated using our simplified model are in close agreement with those estimated using finite element simulations, and thereby confirm the utility of the simplified model. For MMRAs that have been developed for ULF communication applications \cite{Rhinithaa, UCLA, Gong2018, 8304977, 8917925,8457248}, we find that the eddy current loss is appreciable. Notably, it sets the lower bound on the best case power efficiency for these types of magneto-mechanical transmitters, even if the suspension induced losses are considered negligible. Even so, this work supports the argument that MMRA-based ULF transmitters can be orders-of-magnitude more power-efficient and compact than existing ULF transmitters based on antennas or induction coils \cite{alma99955070279905899,Rhinithaa,UCLA}.

\appendices
\section{Torque between Two Magnets of Arbitrary Geometry}
\label{sec:Magnetic_Torque}
\renewcommand{\theequation}{A\arabic{equation}}
\renewcommand{\thefigure}{A\arabic{figure}}
\setcounter{equation}{0}
\setcounter{figure}{0}

\begin{figure}[t]
    \centering
       \includegraphics[width=\linewidth]{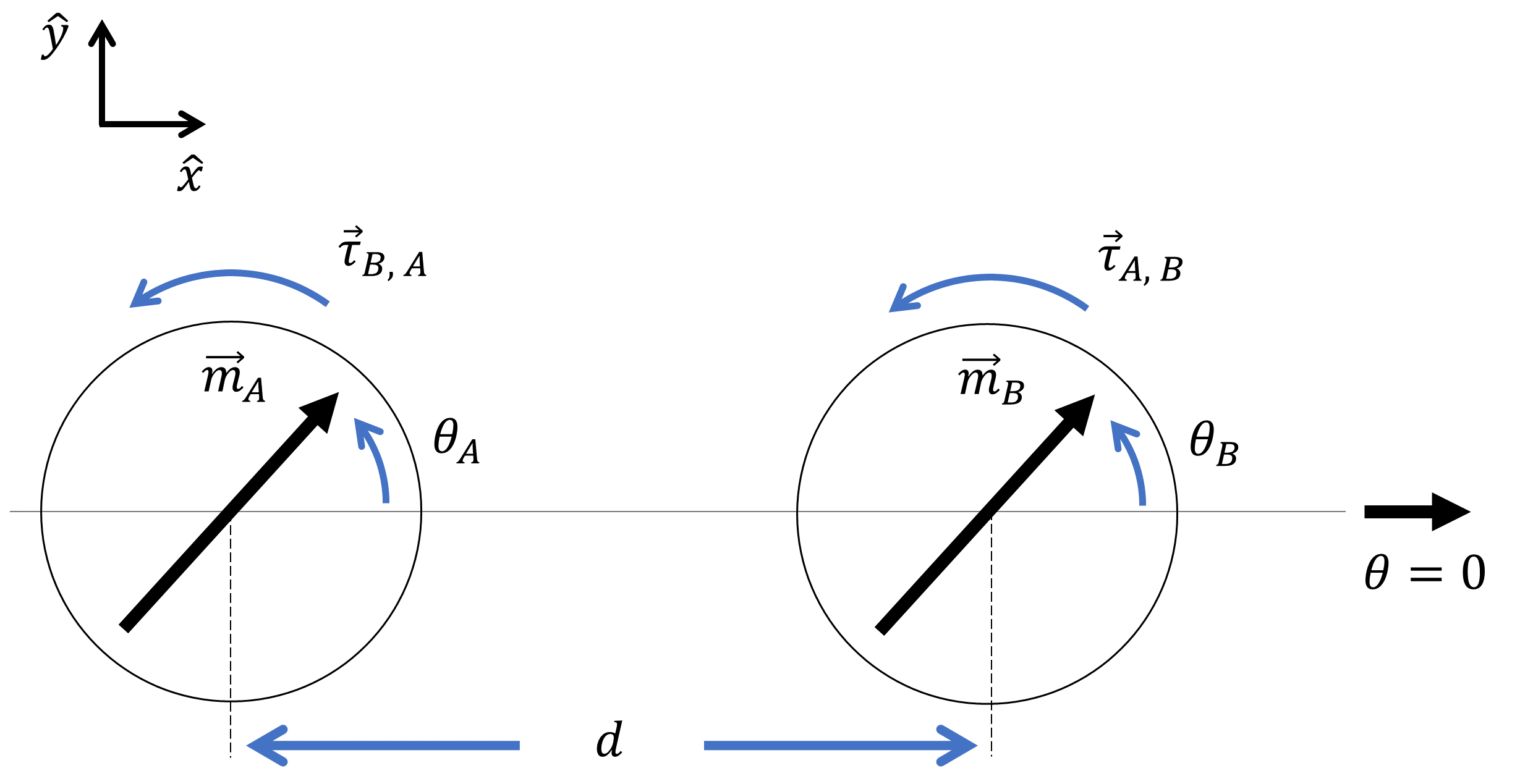}
  \caption{Schematic of the magnetic interaction between two magnets.}
  \label{Appendix_schematic} 
\end{figure}

In this appendix, we derive an analytical expression for the mutual torque between two permanent magnets. The position vector from the centroid of magnet A to the centroid of magnet B is defined as $\vec{r}_{AB} = d\,\hat{x}$. The assumption is made that both magnets have uniform magnetization only in the $xy$ plane. The angles of the magnetization vector with respect to the $\hat{x}$ axis are defined as $\theta_A$ and $\theta_B$ as shown in Fig.~\ref{Appendix_schematic}. 

The magnetic torque can be most simply analytically obtained using the dipole approximation. Here, the magnets are considered as point dipoles at their centroids, and therefore the magnetic torque generated by magnet A on magnet B can be expressed as \cite{Landecker1999}
\begin{align}\label{Appendix_tau_dipole}
    \nonumber \vec{\tau}_{BA} & = \frac{\mu_0 m_A m_B}{4\pi |\vec{r}_{AB}|^3}\left [ 3(\hat{m}_A\cdot\hat{r}_{AB})(\hat{m}_B\times\hat{r}_{AB}) + (\hat{m}_A \times \hat{m}_B)\right] \\ 
    & = - \frac{\mu_0 m_A m_B}{4\pi |d|^3} \left [ 2\,\cos(\theta_A)\sin(\theta_B) + \cos(\theta_B)\sin(\theta_A)\right] \hat{z}
\end{align}
where $\mu_0 = 4\pi\times10^{-7}$ H/m is the vacuum permeability, and $\vec{m}_A = m_A \hat{m}_A$ and $\vec{m}_B = m_B \hat{m}_B$ are the magnetic dipole moments. This dipole approximation method, however, is not accurate when the distance between the magnets is short. Therefore, we develop a new analytical torque expression that can account for the near field effects of magnets and provide a much accurate torque estimate.

The magnetic flux density experienced by an infinitesimal volume element of magnet B due to an infinitesimal element of magnet A is given by
\begin{equation}\label{Appendix_dB}
    d\vec{B}_{AB} = dB_{AB,\,x}\hat{x} + dB_{AB,\,y}\hat{y} + dB_{AB,\,z}\hat{z}
\end{equation}
where 
\begin{subequations}\label{Appendix_dB_xyz}
\begin{align}\label{Appendix_dBx}
\nonumber dB_{AB,\,x} = \frac{B_{r,\,A}}{4\pi} & \left( \frac{3\left[r_{xx^{\prime}}\cos(\theta_A) + r_{yy^{\prime}}\sin(\theta_A) \right]}{\left( r_{xx^{\prime}}^2 + r_{yy^{\prime}}^2 +r_{zz^{\prime}}^2\right)^{(5/2)}} r_{xx^{\prime}} \right. \\
& \quad \left. - \frac{\cos(\theta_A)}{\left( r_{xx^{\prime}}^2 + r_{yy^{\prime}}^2 +r_{zz^{\prime}}^2\right)^{(3/2)}} \right) dV_A
\end{align}
\begin{align}\label{Appendix_dBy}
\nonumber dB_{AB,\,y} = \frac{B_{r,\,A}}{4\pi} & \left( \frac{3\left[r_{xx^{\prime}}\cos(\theta_A) + r_{yy^{\prime}}\sin(\theta_A) \right]}{\left( r_{xx^{\prime}}^2 + r_{yy^{\prime}}^2 +r_{zz^{\prime}}^2\right)^{(5/2)}} r_{yy^{\prime}} \right. \\
  & \quad \left. - \frac{\sin(\theta_A)}{\left( r_{xx^{\prime}}^2 + r_{yy^{\prime}}^2 +r_{zz^{\prime}}^2\right)^{(3/2)}} \right) dV_A
\end{align}
\begin{align}\label{Appendix_dBz}
dB_{AB,\,z} = \frac{B_{r,\,A}}{4\pi} & \left( \frac{3\left[r_{xx^{\prime}}\cos(\theta_A) + r_{yy^{\prime}}\sin(\theta_A) \right]}{\left( r_{xx^{\prime}}^2 + r_{yy^{\prime}}^2 +r_{zz^{\prime}}^2\right)^{(5/2)}} r_{zz^{\prime}} \right) dV_A
\end{align}
\end{subequations}
where $B_{r,\,A}$ is the residual flux density of magnet A, and $dV_A$ is the volume element. Further,
\begin{subequations}\label{Appendix_r}
\begin{align}\label{Appendix_rxx}
r_{xx^{\prime}} = d + x_B - x_A
\end{align}
\begin{align}\label{Appendix_ryy}
r_{yy^{\prime}} = y_B - y_A
\end{align}
\begin{align}\label{Appendix_rzz}
r_{zz^{\prime}} = z_B - z_A
\end{align}
\end{subequations}
where $x_A\hat{x} + y_A\hat{y} + z_A\hat{z}$ is the position vector from the centroid of magnet A to the volume element of magnet A that is under consideration, and $x_B\hat{x} + y_B\hat{y} + z_B\hat{z}$ is the position vector from the centroid of magnet B to the volume element of magnet B that is under consideration. 
The average magnetic flux density produced by magnet A on magnet B in $\hat{x}$-direction and $\hat{y}$ direction can be evaluated as
\begin{equation}\label{Appendix_Bx}
    B_{AB,\,x} = b^{(1)}\, \cos(\theta_A) + b^{(2)}\, \sin(\theta_A) 
\end{equation}
\begin{equation}\label{Appendix_By}
    B_{AB,\,y} = b^{(2)} \, \cos(\theta_A) + b^{(3)}\, \sin(\theta_A) 
\end{equation}
where $b^{(1)}$, $b^{(2)}$, and $b^{(3)}$ are functions of $\theta_A$ and $\theta_B$ given by
\begin{subequations}\label{Appendix_b1_to_b3}
\begin{align}\label{Appendix_b1}
b^{(1)}(\theta_A,\,\theta_B)  = \frac{1}{V_B} \int_{V_B} \int_{V_A} (C_1 - C_2) \, dV_A \, dV_B
\end{align}
\begin{align}\label{Appendix_b2}
b^{(2)}(\theta_A,\,\theta_B)  = \frac{1}{V_B} \int_{V_B} \int_{V_A} C_3 \, dV_A \, dV_B
\end{align}
\begin{align}\label{Appendix_b3}
b^{(3)}(\theta_A,\,\theta_B)  = \frac{1}{V_B} \int_{V_B} \int_{V_A} (C_4 - C_2) \, dV_A \, dV_B
\end{align}
\end{subequations}
where 
\begin{subequations}\label{Appendix_C1_to_C4}
\begin{align}\label{Appendix_C1}
C_1 = \frac{B_{r,\,A}}{4\pi} \frac{3r_{xx^{\prime}}^2}{\left( r_{xx^{\prime}}^2 + r_{yy^{\prime}}^2 + r_{zz^{\prime}}^2\right)^{(5/2)}}
\end{align}
\begin{align}\label{Appendix_C2}
C_2 = \frac{B_{r,\,A}}{4\pi} \frac{1}{\left( r_{xx^{\prime}}^2 + r_{yy^{\prime}}^2 + r_{zz^{\prime}}^2\right)^{(3/2)}}
\end{align}
\begin{align}\label{Appendix_C3}
C_3 = \frac{B_{r,\,A}}{4\pi} \frac{3r_{xx^{\prime}}r_{yy^{\prime}}}{\left( r_{xx^{\prime}}^2 + r_{yy^{\prime}}^2 + r_{zz^{\prime}}^2\right)^{(5/2)}}
\end{align}
\begin{align}\label{Appendix_C4}
C_4 = \frac{B_{r,\,A}}{4\pi} \frac{3r_{yy^{\prime}}^2}{\left( r_{xx^{\prime}}^2 + r_{yy^{\prime}}^2 + r_{zz^{\prime}}^2\right)^{(5/2)}} ~.
\end{align}
\end{subequations}
The magnetic torque acting on the volume element of magnet B produced by the volume element of magnet A along the $\hat{z}$ axis is given by
\begin{align}\label{Appendix_dtau}
    \nonumber d{\vec{\tau}}_{BA} & = \frac{B_{r,\,B}}{\mu_0}\left[ \cos(\theta_B)\hat{x} + \sin(\theta_B)\hat{y}\right] dV_B \times d\vec{B}_{AB} \\
    \nonumber & = \frac{B_{r,\,B}}{\mu_0}\left[ C_3 \, \cos(\theta_{A} + \theta_{B}) \right. \\
    \nonumber & \quad \quad \quad \quad + (C_4 - C_2) \, \sin(\theta_{A})\cos(\theta_{B}) \\ 
    & \left. \quad \quad \quad \quad + (C_2 - C_1) \, \cos(\theta_{A})\sin(\theta_{B}) \right] \, dV_A \, dV_B \hat{z}
\end{align}
where $B_{r,\,B}$ is the residual flux density of magnet B. Therefore, the total magnetic torque acting on magnet B by magnet A can be evaluated as
\begin{align}\label{Appendix_tau}
    \nonumber \vec{\tau}_{BA} = - &\left[ \kappa^{(1)} \, \cos(\theta_{A}+\theta_{B}) + \kappa^{(2)} \, \sin(\theta_{A})\cos(\theta_{B}) \right. \\
    & \left. + \kappa^{(3)} \, \cos(\theta_{A})\sin(\theta_{B}) \right] \hat{z}
\end{align}
where $\kappa^{(1)}$, $\kappa^{(2)}$, and $\kappa^{(3)}$ are functions of $\theta_A$ and $\theta_B$ given by
\begin{subequations}\label{Appendix_k1_to_k3}
\begin{align}\label{Appendix_k1}
\kappa^{(1)}(\theta_{A},\theta_{B}) = -\frac{B_{r,\,B}}{\mu_0} \int_{V_B} \int_{V_A} C_3 \, dV_A \, dV_B
\end{align}
\begin{align}\label{Appendix_k2}
\kappa^{(2)}(\theta_{A},\theta_{B}) = -\frac{B_{r,\,B}}{\mu_0} \int_{V_B} \int_{V_A} ( C_4 - C_2) \, dV_A \, dV_B
\end{align}
\begin{align}\label{Appendix_k3}
\kappa^{(3)}(\theta_{A},\theta_{B}) = -\frac{B_{r,\,B}}{\mu_0} \int_{V_B} \int_{V_A} (C_2 - C_1) \, dV_A \, dV_B ~.
\end{align}
\end{subequations}

Parameters $b^{(1)}$, $b^{(2)}$, $b^{(3)}$, and $\kappa^{(1)}$, $\kappa^{(2)}$, $\kappa^{(3)}$ are functions of $\theta_A$ and $\theta_B$, since as $\theta_A$ and $\theta_B$ change, the volume of integration will change accordingly. Therefore, when geometries of magnet A and magnet B have circular symmetry around the $\hat{z}$ axis that goes through their centroids, the parameters $b^{(1)}$, $b^{(2)}$, $b^{(3)}$, and $\kappa^{(1)}$, $\kappa^{(2)}$, $\kappa^{(3)}$ will become constants and will be independent from $\theta_A$ and $\theta_B$. Additionally, in the special case where magnet A and magnet B are symmetric about the $\hat{x}$ axis, $b^{(2)}$ and $\kappa^{(1)}$ will become zero.

\section{Eddy current loss coefficients for cylindrical and cuboidal conducting objects}
\label{sec:Eddy_current_loss_coeff}

\renewcommand{\theequation}{B\arabic{equation}}
\renewcommand{\thefigure}{B\arabic{figure}}
\renewcommand{\thetable}{B\arabic{table}}
    
\setcounter{equation}{0}
\setcounter{figure}{0}
\setcounter{table}{0}

\begin{figure}[ht]
    \centering
  {\includegraphics[width=\linewidth]{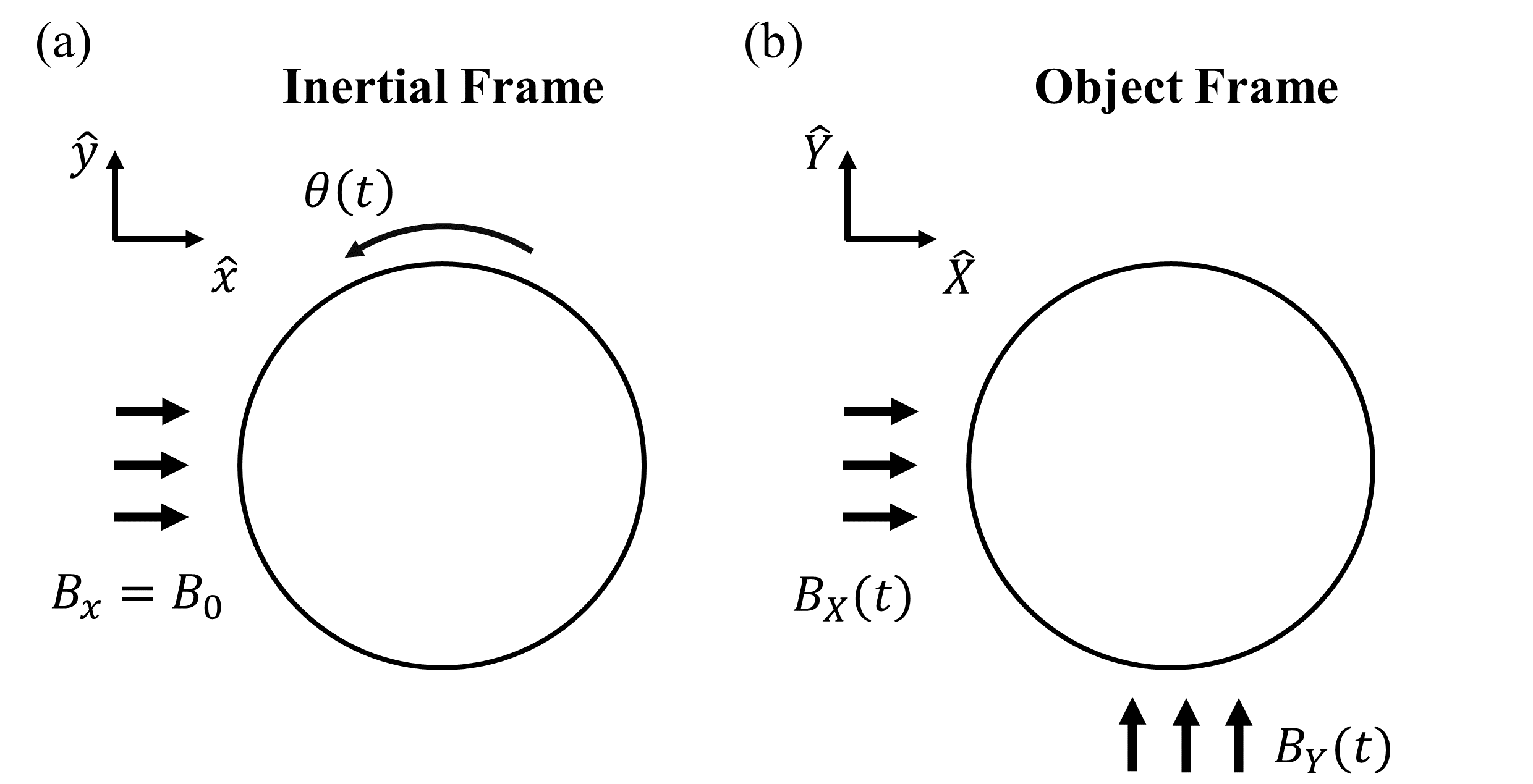}}
\caption{\textbf{(a)} Schematic of a cylindrical conductor rotating in a uniform magnetic field along the transverse direction in the inertial frame. $B_0$ is the magnitude of the external magnetic flux density. $\theta(t)$ is the instantaneous angular position of the conductor. \textbf{(b)} Schematic of the conductor in the object frame.}
  \label{AppendixB_model} 
\end{figure}

The eddy current loss for a cylindrical conductor rotating in a uniform magnetic field transverse to its axis [Fig.~\ref{AppendixB_model}(a)] has been analytically derived in \cite{Aubert2012}, and can be expressed as 
\begin{align}\label{Appendix_B_P_rot}
    P_e = \frac{128}{\pi^3} \sigma a^2 b^3 B_0^2 \omega^2 \sum_{i} \frac{1}{(2i+1)^4} \frac{J_2(ma)}{J_0(ma) + J_2(ma)}
\end{align} 
where $\sigma$ is the conductivity of the conductor, $a$ is the radius, $b$ is half of the length, $i = 0, 1, ...$, $m = \frac{(2i+1)\pi}{2b}$, $J_0$ and $J_2$ are the Bessel functions of the first kind, $B_0$ is the magnitude of the magnetic flux density, and $\omega$ is the constant angular velocity of the conductor. We can study the eddy current loss in the object frame [Fig.~\ref{AppendixB_model}(b)], and using (\ref{peddy_XY}) we can get
\begin{align}\label{Appendix_B_P_obj}
    P_e = P_{e,\,X} + P_{e,\,Y} = \frac{A_e^2}{R_e}(\dot{B}_X^2 + \dot{B}_Y^2)
\end{align} 
where $\frac{A_e^2}{R_e}$ is the eddy current loss coefficient of the cylindrical conductor along the transverse direction, and using (\ref{B_X}) and (\ref{B_Y}), $B_X$ and $B_Y$ can be expressed as
\begin{align}
    B_{X} 
        = & B_0\,\textrm{cos}(\omega t) \label{AppendixB_B_X}\\
    B_{Y} 
        = & - B_0\,\textrm{sin}(\omega t) \label{AppendixB_B_Y} ~.
\end{align}
Substituting (\ref{AppendixB_B_X}) and (\ref{AppendixB_B_Y}) into (\ref{Appendix_B_P_obj}), we can get the eddy current loss as
\begin{align}\label{Appendix_B_P_rot_2}
    P_e = \frac{A_e^2}{R_e}B_0^2 \omega_r^2 ~.
\end{align} 
Using (\ref{Appendix_B_P_rot}) and (\ref{Appendix_B_P_rot_2}), we can get the eddy current loss coefficient of a cylindrical conductor along the transverse direction as
\begin{align}\label{Appendix_prefactor}
    \frac{A_e^2}{R_e} = \frac{128}{\pi^3} \sigma a^2 b^3 \sum_{i} \frac{1}{(2i+1)^4} \frac{J_2(ma)}{J_0(ma) + J_2(ma)} ~.
\end{align}

The amplitude of the eddy current loss for a cuboidal conductor in an oscillating magnetic field $B_0\,\textrm{cos}(\omega t)$ (which points purely along a surface normal) is given by the following well known expression \cite{767162}:
\begin{align}\label{Appendix_eddy_current_loss_cub}
    P_e = \frac{1}{12}\sigma\omega^2d_t^2B_0^2F_{sk} V_\textrm{cub}
\end{align}
where $d_t$ is the thickness of the cuboid along the direction corresponding to the surface normal, $V_\textrm{cub}$ is the total volume of the cuboidal conductor, and $F_{sk}$ is the skin effect factor, which can be reasonably approximated as 1 in the ULF range \cite{767162}. Using (\ref{Appendix_eddy_current_loss_cub}) and (\ref{Appendix_B_P_rot_2}), we can get the eddy current loss coefficient of a cuboidal conductor as
\begin{align}\label{Appendix_Effective_A_stator}
    \frac{A_e^2}{R_e} = \frac{\sigma d_t^2}{12} V_\textrm{cub} ~.
\end{align}

\section{Values of Key Parameters for Tested MMRA Prototypes}
\label{sec:Values_of_Key_Parameters}

\renewcommand{\theequation}{B\arabic{equation}}
\renewcommand{\thefigure}{B\arabic{figure}}

\setcounter{equation}{0}
\setcounter{figure}{0}

Table \ref{Dimension_and_parameter} lists the key parameters for the MMRA prototypes that we have tested, as described in Fig.~\ref{Devices}.

\setcounter{table}{0}
    \renewcommand{\thetable}{C\arabic{table}}
    
 \begin{table}[!htp]
	\renewcommand{\arraystretch}{1.3}
	\caption{Values of key parameters}
	\centering
	\label{Dimension_and_parameter}
	\resizebox{\columnwidth}{!}{
		\begin{tabular}{c c c}
			\hline\hline \\[-3mm]
Parameters         &Single-rotor       &Multi-rotor      \\
~                  &MMRA               &MMRA      \\
\hline
Rotor magnet remanence            & 1.338 {T}        & 1.349 {T}           \\ 
Rotor length                & 2.3125 {in}      & 2.8125 {in}           \\ 
Rotor diameter                    & 4 {mm}           & 3.5 {mm}          \\ 
Rotor moment of inertia           & 13.75 {$\textrm{mm}^2\textrm{g}$}   & 17.50 {$\textrm{mm}^2\textrm{g}$}          \\ 
Suspension torsional stiffness       & 2.2 {$\textrm{N\,mm}/\textrm{rad}$}  & 3.9 {$\textrm{N\,mm}/\textrm{rad}$}         \\ 
Stator magnet remanence           & 1.355 {T}        & 1.164 {T}         \\ 
Stator length   & 3 {in}           & 3 {in}          \\ 
Stator width    & 0.5 {in}         & 0.125 {in}           \\ 
Stator height  & 0.5 {in}         & 0.5 {in}          \\ 
Conductivity of magnets& 0.625 {$\textrm{MS}/\textrm{m}$}    & 0.625 {$\textrm{MS}/\textrm{m}$}           \\ 
Center distance between rotors    & {$\textrm{N}/\textrm{A}$}            & 6.5 {mm}           \\ 
Center distance between stator    & 21.85 {mm}               & 9.68 {mm}            \\ 
and the nearest rotor             &        &           \\ 
Coupling coefficient              & 0.68 {$\textrm{N\,mm}/\textrm{A}$}              & 0.84 {$\textrm{N\,mm}/\textrm{A}$} \\
\hline\hline
		\end{tabular}
	}
\end{table}

\section{Shape of the In-Phase Mode of the Multi-Rotor MMRA in the Nonlinear Regime}
\label{sec:Nonlinear_Mode_Shape}

\renewcommand{\theequation}{D\arabic{equation}}
\renewcommand{\thefigure}{D\arabic{figure}}
\setcounter{equation}{0}
\setcounter{figure}{0}

\begin{figure}[ht]
    \centering
    {\includegraphics[width=\linewidth]{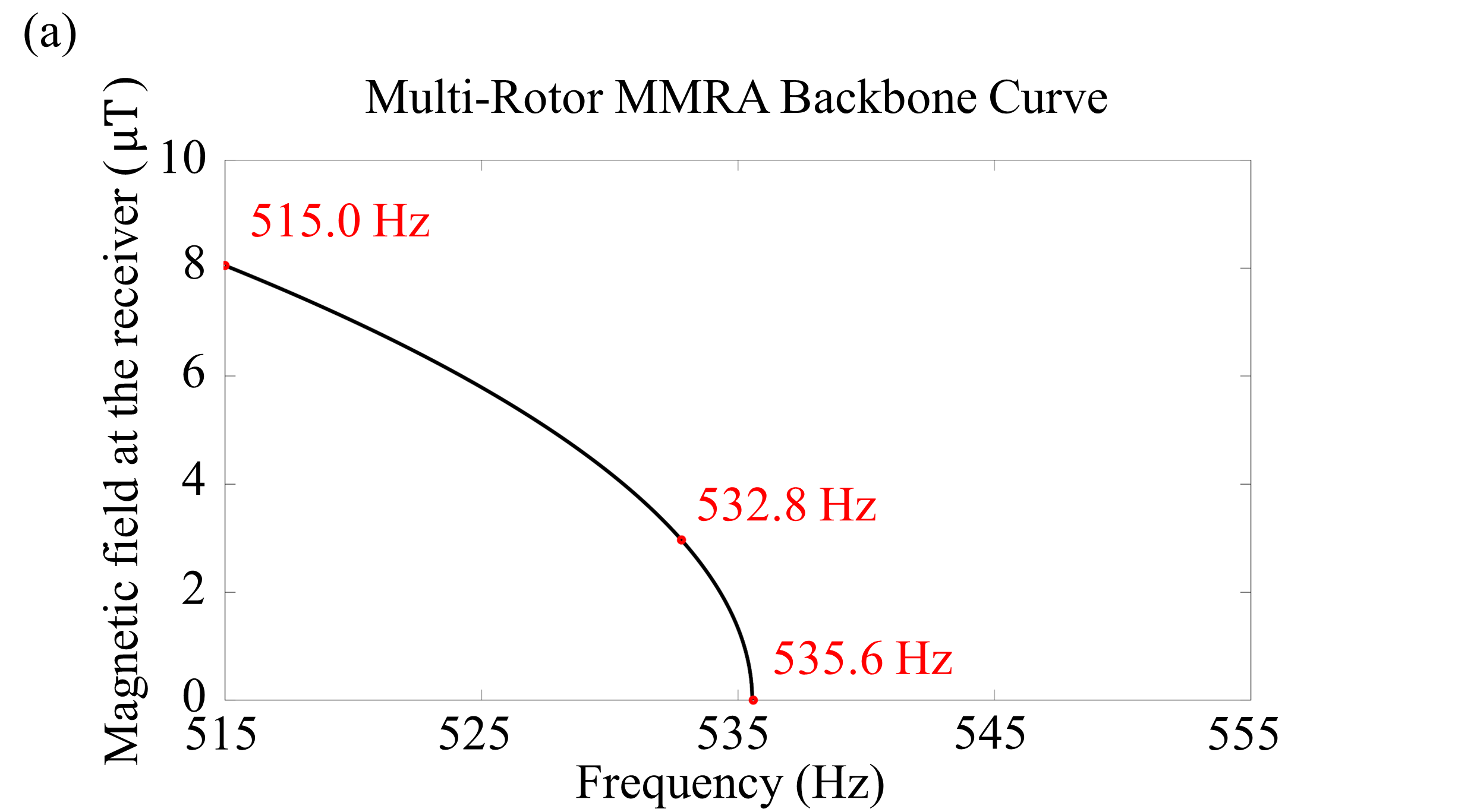}}
    \\
    {\includegraphics[width=\linewidth]{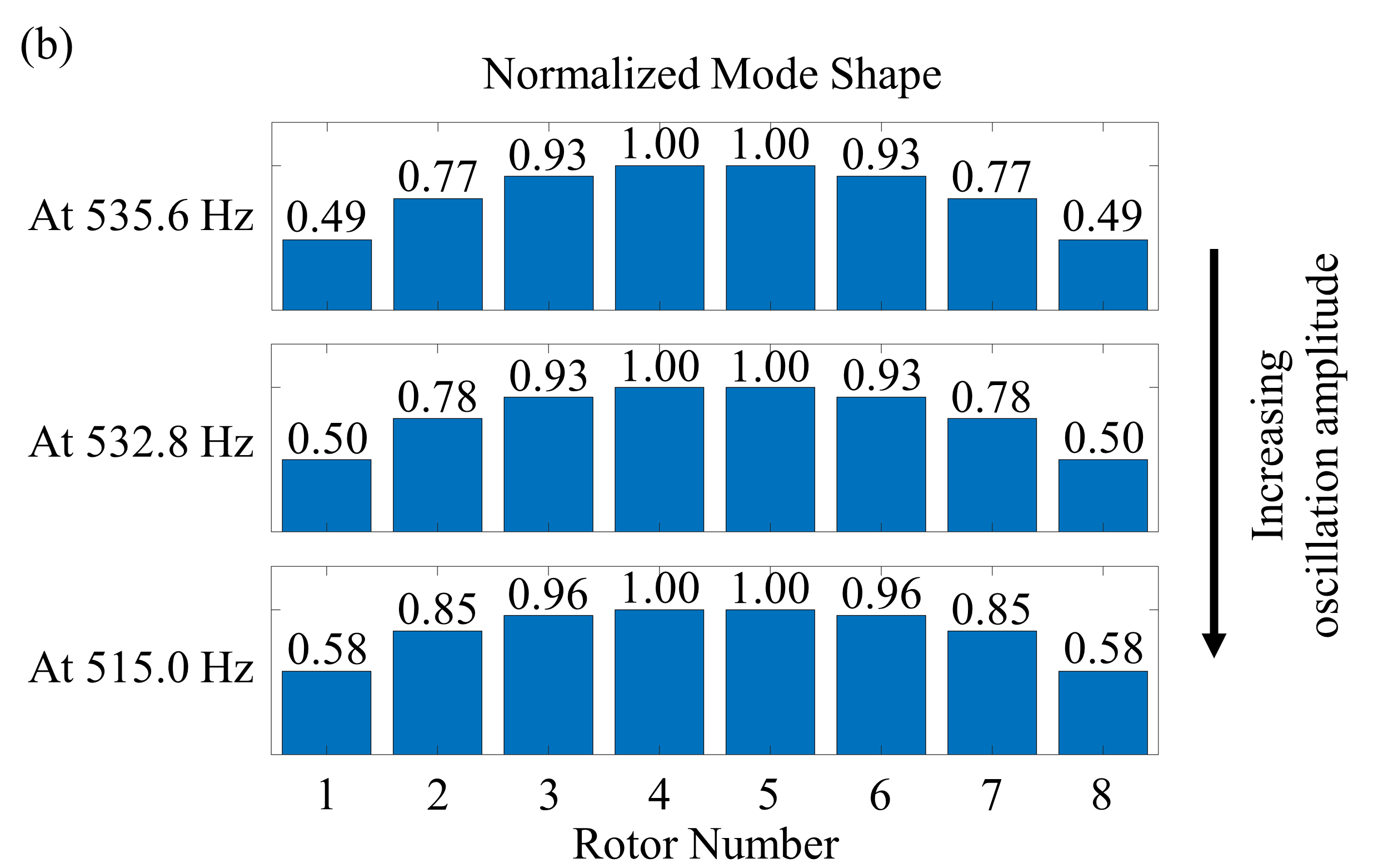}}
    \caption{\textbf{(a)} Amplitude-frequency nonlinearity induced backbone curve for the multi-rotor MMRA. \textbf{(b)} Normalized mode shape at three different resonant frequencies (highlighted in (a)).}
    \label{mode_shape_appendix} 
\end{figure}

The oscillation mode shape for the in-phase mode of the multi-rotor MMRA is a function of the drive amplitude because of the nonlinearity intrinsic to the magneto-mechanical coupling. This mode shape can be evaluated with the help of (\ref{eom_omega}). In Fig.~{\ref{mode_shape_appendix}}(a), we present the amplitude-frequency effect backbone curve for the tested multi-rotor MMRA, as was previously evaluated in Fig.~\ref{Dyanmic_results}. Fig.~{\ref{mode_shape_appendix}}(b) presents the mode shape (normalized to maximum rotor amplitude) for three specific resonant frequencies along the curve. As shown in the figure, the mode shape distributes a little more evenly across the resonators as the oscillation amplitude increases, which can be beneficial for reducing power consumption for a given field level \cite{Rhinithaa}.

\section*{Acknowledgment}
This work was sponsored by the Defense Advanced Research Projects Agency grant HR0011-17-2-0057 under the AMEBA program, by the U.S. National Science Foundation Emerging Frontiers in Research and Innovation program and by the Office of Naval Research Director of Research Early Career Grant (grant N00014-17-1-2209). We additionally extend thanks to Dr. Jiho Noh, Mr. Chengzhang Li, and Mr. Gengming Liu for their insights.

\bibliographystyle{ieeetr}
\bibliography{main}

\end{document}